# Equivalence Principles, Spacetime Structure and the Cosmic Connection

Wei-Tou Ni

*Center for Gravitation and Cosmology,*
*Department of Physics, National Tsing Hua University,*
*No. 101, Kuang Fu II Rd., Hsinchu, ROC 30013*
weitou@gmail.com

After reviewing the meaning of various equivalence principles and the structure of electrodynamics, we give a fairly detailed account of the construction of the light cone and a core metric from the equivalence principle for the photon (no birefringence, no polarization rotation and no amplification/attenuation in propagation) in the framework of linear electrodynamics using cosmic connections/observations as empirical support. The cosmic nonbirefringent propagation of photons independent of energy and polarization verifies the Galileo Equivalence Principle [Universality of Propagation] for photons/electromagnetic wave packets in spacetime. This nonbirefringence constrains the spacetime constitutive tensor to high precision to a core metric form with an axion degree and a dilaton degree of freedom. Thus comes the metric with axion and dilation. Constraints on axion and dilaton from astrophysical/cosmic propagation are reviewed. Eötvös-type experiments, Hughes-Drever-type experiments, redshift experiments then constrain and tie this core metric to agree with the matter metric, and hence a unique physical metric and universality of metrology. We summarize these experiments and review how the Galileo equivalence principle constrains the Einstein Equivalence Principle (EEP) theoretically. In local physics this physcycal metric gives the Lorentz/Poincaré covariance. Understanding that the metric and EEP come from the vacuum as a medium of electrodynamics in the linear regime, efforts to actively look for potential effects beyond this linear scheme are warranted. We emphasize the importance of doing Eötvös-type experiments or other type experiments using polarized bodies/polarized particles. We review the theoretical progress on the issue of gyrogravitational ratio for fundamental particles and update the experimental progress on the measurements of possible long range/intermediate range spin-spin, spin-monopole and spin-cosmos interactions.

## 1. Introduction

In the genesis of general relativity, there are two important cornerstones: the Einstein Equivalence Principle (EEP) and the metric as the dynamic quantity of gravitation (See, e.g., Ref. [1]). With research activities on cosmology thriving, people have been looking actively for alternative theories of gravity again for more than thirty years. Recent theoretical studies include scalars, pseudoscalars, vectors, metrics, bimetrics, strings, loops, etc. as dynamic quantities of gravity. It is the aim of this review to look for the foundations of gravity and general relativity, especially from an empirical point of view.

Relativity sprang out from Maxwell-Lorentz theory of electromagnetism. Maxwell equations in Gaussian units are

$$\nabla \cdot \boldsymbol{D} = 4\pi \rho, \tag{1a}$$
$$\nabla \times \boldsymbol{H} - \partial \boldsymbol{D}/\partial t = 4\pi \boldsymbol{J}, \tag{1b}$$
$$\nabla \cdot \boldsymbol{B} = 0, \tag{1c}$$
$$\nabla \times \boldsymbol{E} + \partial \boldsymbol{B}/\partial t = 0, \tag{1d}$$

where $\boldsymbol{D}$ is the displacement, $\boldsymbol{H}$ the magnetic field, $\boldsymbol{B}$ the magnetic induction, $\boldsymbol{E}$ the electric field, $\rho$ the electric charge density, and $\boldsymbol{J}$ the electric current density. We use units with the nominal light velocity $c$ equal to 1 (See, e.g., Jackson [2], p. 218 (6.28)). With the sources known, from these equations with 8 components we are supposed to be able to solve for the unknown fields $\boldsymbol{D}$, $\boldsymbol{H}$, $\boldsymbol{B}$ and $\boldsymbol{E}$ with 12 degrees of freedom. These equations form an under determined system unless we supplement them with relations. These relations are the constitutive relation between ($\boldsymbol{D}$, $\boldsymbol{H}$) and ($\boldsymbol{E}$, $\boldsymbol{B}$) [or ($\boldsymbol{D}$, $\boldsymbol{B}$) and ($\boldsymbol{E}$, $\boldsymbol{H}$)]:

$$(\boldsymbol{D}, \boldsymbol{H}) = \chi(\boldsymbol{E}, \boldsymbol{B}), \tag{2}$$

where $\chi(\boldsymbol{E}, \boldsymbol{B})$ is a 6-component functional of $\boldsymbol{E}$ and $\boldsymbol{B}$. With the constitutive relation, the unknown degrees of freedom become 6, the Maxwell equations seem to be over determined. Note that if we take the divergence of (1d), by (1c) it is automatically satisfied. Hence (1c) and (1d) (the Faraday tetrad) have only 3 independent equations. If we take the divergence of (1b), by (1a) it becomes the continuity equation

$$\nabla \cdot \boldsymbol{J} + \partial \rho/\partial t = 0, \tag{3}$$

a constraint equation on sources. Hence, (1a) and (1b) (the Ampère-Maxwell tetrad) have only 3 independent equations also. To form a complete system of equations, we need equations governing the action of the electric field and magnetic induction on the charge and current. Lorentz force law provides this link and completes the system:

$$\boldsymbol{F} = m \, d\boldsymbol{v}/dt = q \, (\boldsymbol{E} + \boldsymbol{v} \times \boldsymbol{B}), \tag{4}$$

where $\boldsymbol{v}$ is the velocity of the charge and $\boldsymbol{F}$ is the force on it due to electric field and magnetic induction.

In 1908, Minkowski [3,4] put the Maxwell equations into geometric form in four-dimensional spacetime with Lorentz covariance using Cartesian coordinates $x$, $y$, $z$ and imaginary time $it$ and numbering them as $x_1 \equiv x$, $x_2 \equiv y$, $x_3 \equiv z$ and $x_4 \equiv it$. Minkowski defined the 4-dim excitation $^{(\text{Mink})}f$ and the 4-dim field strength $^{(\text{Mink})}F$ as

$$^{(\text{Mink})}f \equiv (^{(\text{Mink})}f_{hk}) \equiv \begin{pmatrix} 0 & H_z & -H_y & -iD_x \\ -H_z & 0 & H_x & -iD_y \\ H_y & -H_x & 0 & -iD_z \\ iD_y & iD_y & iD_z & 0 \end{pmatrix}, \tag{5a}$$

$$^{(\text{Mink})}F \equiv (^{(\text{Mink})}F_{hk}) \equiv \begin{pmatrix} 0 & B_z & -B_y & -iE_x \\ -B_z & 0 & B_x & -iE_y \\ B_y & -B_x & 0 & -iE_z \\ iE_y & iE_y & iE_z & 0 \end{pmatrix}. \tag{5b}$$

In terms of these quantities, Minkowski put the Maxwell equation into the 4-dim covariant form:

$$^{(\text{Mink})}f_{hk,h} = -s_k, \tag{6a}$$
$$^{(\text{Mink})}F^*{}_{hk,h} = 0, \tag{6b}$$

with

$$F^*{}_{hk} \equiv (1/2)\, \varrho_{hklm}\, F_{lm}, \text{ and } s_k \text{ the 4-current.} \tag{6c}$$

Here $\varrho_{hklm} = \pm 1, 0$ is the totally antisymmetric Levi-Civita symbol with $\varrho_{1234} = +1$. The equations (6a,b) are covariant in the sense that for the linear transformations with constant coefficients that leave the form

$$x_h x_h \equiv (x_1)^2 + (x_2)^2 + (x_3)^2 + (x_4)^2 \tag{7}$$

invariant, the Maxwell equations in the form (6a,b) are covariant with the 4-dim excitation $f_{\text{Minkowski}}$ and the 4-dim field strength $F_{\text{Minkowski}}$ transforming as 4-dim covariant tensors (covariant *V*-six-vectors).

Bateman [5] used time coordinate $t$ instead of $x_4$, and considered transformations that leave the invariance of the differential (form) equation:

$$(dx)^2 + (dy)^2 + (dz)^2 - (dt)^2 = 0. \tag{8}$$

Hence, he also included conformal transformations in addition to Lorentz transformations and made one step toward general coordinate invariance. With indefinite metric, one has to distinguish covariant and contravariant tensors and indices. Aware of this, one can readily put Maxwell equations into covariant form without using imaginary time.

In terms of field strength $F_{kl}$ (**E**, **B**) and excitation (density) $H^{ij}$ (**D**, **H**):

$$F_{kl} = \begin{pmatrix} 0 & E_1 & E_2 & E_3 \\ -E_1 & 0 & -B_3 & B_2 \\ -E_2 & B_3 & 0 & -B_1 \\ -E_3 & -B_2 & B_1 & 0 \end{pmatrix}, \tag{9a}$$

$$H^{ij} = \begin{pmatrix} 0 & -D_1 & -D_2 & -D_3 \\ D_1 & 0 & -H_3 & H_2 \\ D_2 & H_3 & 0 & -H_1 \\ D_3 & -H_2 & H_1 & 0 \end{pmatrix}, \tag{9b}$$

Maxwell equations can be expressed as

$$H^{ij}{}_{,j} = -4\pi J^i, \qquad (10a)$$
$$e^{ijkl} F_{jk,l} = 0, \qquad (10b)$$

with the constitutive relation (2) between the excitation and the field in the form:

$$H^{ij} = \chi^{ij}(F_{kl}), \qquad (11)$$

where $J^k$ is the charge 4-current density $(\rho, \mathbf{J})$ and $e^{ijkl}$ the completely anti-symmetric tensor density (Levi-Civita symbol) with $e^{0123} = 1$ (See, e. g., Hehl and Obukhov [6]). Here is $\chi^{ij}(F_{kl})$ is a functional with 6 independent degrees of freedom. For medium with a local linear response or in the linear local approximation, (11) reduced to

$$H^{ij} = \chi^{ijkl} F_{kl}, \qquad (12)$$

with $\chi^{ijkl}$ the (linear) constitutive tensor density [6-10]. For isotropic dielectric and isotropic permeable medium, the constitutive tensor density has 2 degrees of freedom; for anisotropic dielectric and anisotropic permeable medium, the constitutive tensor density has 12 degrees of freedom; for general linear local medium (with magnetoelectric response), the constitutive tensor has 21 degrees of freedom.

Introducing the metric $g_{ij}$ as gravitational potential in 1913 [11] and versed in general (coordinate-)covariant formalism in 1914 [12], Einstein put the Maxwell equations in general covariant form ($\mathcal{F}^{ij} = H^{ij}$ in our notation) [12]:

$$\mathcal{F}^{ij}{}_{,j} = -4\pi J^i, \qquad (13a)$$
$$F_{ij,k} + F_{jk,i} + F_{ki,j} = 0. \qquad (13b)$$

Shortly after Einstein constructed general relativity, Einstein noticed that the Maxwell equations can be formulated in a form independent of the metric gravitational potential in 1916 [13]. Einstein introduced the covariant *V*-six-vector Equations (13a) and (13b) which are independent of metric gravitational potential. Only the constitutive tensor density $\chi^{ijkl}$ is dependent on the metric gravitational potential:

$$\mathcal{F}^{ij} = (-g)^{1/2} g^{ik} g^{jl} F_{kl}. \qquad (14)$$

Noticing Einstein's $\mathcal{F}^{ij}$ is our $H^{ij}$ and putting (14) in the form of (12), we have

$$\chi^{ijkl} = (-g)^{1/2}[(1/2)g^{ik} g^{jl} - (1/2)g^{il} g^{kj}]. \qquad (15)$$

In local inertial frame the metric-induced constitutive tensor (15) is reduced to special-relativitivistic Minkowski form:

$$\chi^{ijkl} = (-g)^{1/2}[(1/2)\eta^{ik} \eta^{jl} - (1/2)\eta^{il} \eta^{kj}] + O(x^i x^j), \qquad (16)$$

which is dictated by the Einstein equivalence principle.

In macroscopic medium, the constitutive tensor gives the medium-coupling to electromagnetism; it depends on the (thermodynamic) state of the medium and in turn depends on temperature, pressure etc. In gravity, the constitutive tensor gives the gravity-coupling to electromagnetism; it depends on the gravitational field(s) and in turn depends on the matter distribution and its state.

In gravity, a fundamental issue is how to arrive at the metric from the constitutive tensor through experiments and observations. That is, how to build the metric empirically and test the EEP thoroughly. Are there other degrees of freedom to be explored?

Since ordinary energy compared to Planck energy is very small, in this situation we can assume that the gravitational (or spacetime) constitutive tensor is a linear and local function of gravitational field(s), i.e. (12) holds. Since the second half of 1970's, we have started to use the following the Lagrangian density $L$ (= $L_I^{(EM)} + L_I^{(EM-P)}$) with the electromagnetic field Lagrangian $L_I^{(EM)}$ and the field-current interaction Lagrangian $L_I^{(EM-P)}$ given by

$$L_I^{(EM)} = -(1/(16\pi))H^{ij} F_{ij} = -(1/(16\pi))\chi^{ijkl} F_{ij} F_{kl}, \qquad (17a)$$
$$L_I^{(EM-P)} = -A_k J^k, \qquad (17b)$$

for studying this issue [14-16]. Here $\chi^{ijkl} = -\chi^{jikl} = \chi^{klij}$ is a tensor density of the gravitational fields or matter fields to be investigated, $F_{ij} \equiv A_{j,i} - A_{i,j}$ the electromagnetic field strength tensor with $A_i$ the electromagnetic 4-potential and comma denoting partial derivation, and $J^k$ the charge 4-current density. The Maxwell equations (10a,b) or (1a-d) can be derived from this Lagrangian with the relation (12) and (9a,b). Using this $\chi$-framework, we have demonstrated the construction of the light cone core metric from the experiments and observations as in Table 1 [17]. After presenting the meaning of various equivalence principles in section 2 and the structure of premetric electrodynamics in section 3.1, we give a fairly detailed account of the construction of the metric together with constraints on axions, dilatons and skewons from the equivalence principle for photon in the framework of premetric electrodynamics using cosmic observations as empirical support in section 3.2 to section 3.6. Section 3.7 discusses the special case of spacetime/medium with constitutive tensor induced by asymmetric metric and its special role. Section 3.8 addresses the issue of empirical foundation of the closure relation.

In section 4, we review theorems and relations among various equivalence principles using the $\chi$-framework including particles and the corresponding $\chi$-framework for the nonabelian field. In section 5, we discuss the relation of universal metrology and equivalence principles. In section 6, we review theoretical works on the gyrogravitational effects. In section 7, experimental progress on the measurement of long range/intermediate range spin-spin, spin-monopole and spin-cosmos interactions is updated. In section 8, prospects are discussed.

Table 1. Constraints on the spacetime constitutive tensor $\chi^{ijkl}$ and construction of the spacetime structure (metric + axion field $\varphi$ + dilaton field $\psi$) from experiments/observations in skewonless case ($U$: Newtonian gravitational potential). $g_{ij}$ is the particle metric. [17]

| Experiment | Constraints | Accuracy |
|---|---|---|
| Pulsar Signal Propagation | $\chi^{ijkl} \rightarrow \frac{1}{2} (-h)^{1/2}[h^{ik} h^{jl} - h^{il} h^{kj}]\psi + \varphi e^{ijkl}$ | $10^{-16}$ |
| Radio Galaxy Observation | | $10^{-32}$ |
| Gamma Ray Burst (GRB) | | $10^{-38}$ |
| CMB Spectrum Measurement | $\psi \rightarrow 1$ | $8 \times 10^{-4}$ |
| Cosmic Polarization Rotation Experiment | $\varphi - \varphi_0 (\equiv \alpha) \rightarrow 0$ | $|\langle\alpha\rangle| < 0.02$, $\langle(\alpha-\langle\alpha\rangle)^2\rangle^{1/2} < 0.03$ |
| Eötvös-Dicke-Braginsky Experiments | $\psi \rightarrow 1$ $h_{00} \rightarrow g_{00}$ | $10^{-10} U$ $10^{-6} U$ |
| Vessot-Levine Redshift Experiment | $h_{00} \rightarrow g_{00}$ | $1.4 \times 10^{-4} \Delta U$ |
| Hughes-Drever-type Experiments | $h_{\mu\nu} \rightarrow g_{\mu\nu}$ $h_{0\mu} \rightarrow g_{0\nu}$ $h_{00} \rightarrow g_{00}$ | $10^{-24}$ $10^{-19}$ -$10^{-20}$ $10^{-16}$ |

## 2. Meaning of Various Equivalence Principles

Our common understanding and formulation of gravity can be simply described in the following picture: Matter produces gravitational field and gravitational field influences matter. In Newtonian theory of gravity [18], the Galileo Weak Equivalence Principle (WEP I) [19] determines how matter behaves in a gravitational field, and Newton's inverse square law determines how matter produces gravitational field. In a relativistic theory of gravity such as a metric theory, the EEP determines how matter behaves in a gravitational field, and the field equations determine how matter produces gravitational field(s). In Einstein's general relativity, with a suitable choice of the stress-energy tensor, the Einstein equation can imply the Einstein equivalence principle. In non-metric theories of gravity, other versions of equivalence principles may be used. The above situations can be summarized in the following table together with those for electromagnetism.

Table 2. Gravity and Electromagnetism

| Matter | $\xrightarrow{\text{produces}}$ | Gravitational Field(s) | $\xrightarrow{\text{influence(s)}}$ | Matter |
|---|---|---|---|---|
| Newtonian Gravity | Inverse Square Law | | WEP[I] | |
| Relativistic Gravity | Field Equation(s) e.g., Einstein equation | | EEP or substitute | |
| Charges | $\xrightarrow{\text{produce}}$ | Electromagnetic Field | $\xrightarrow{\text{influences}}$ | Charges |
| Electromagnetism | Maxwell Equations | | Lorentz Force Law | |

From Table 2, we see the crucial role played by equivalence principles in the formulation of gravity. In the following, we start with the ancient concepts of inequivalence and discuss meaning of various equivalence principles. This section is an update of Sec. II of Ref. [16].

**2.1.** *Ancient concepts of inequivalence*

From the observations that heavy bodies fall faster than light ones in the air, ancient people, both in the orient and in the west, believe that objects with different constituents behave differently in a gravitational field. We now know that this is due to the inequivalent responses to different buoyancy forces and air resistances.

**2.2.** *Macroscopic equivalence principles*

(i) *Galileo equivalence principle* (WEP I) [19]

Using an inclined plane, Galileo (1564-1642) showed that the distance a falling body travels from rest varies as the square of the time. Therefore, the motion is one of constant acceleration. Moreover, Galileo demonstrated that "the variation of speed in air between balls of gold, lead, copper, porphyry, and other heavy materials is so slight that in a fall of 100 cubits [about 46 meters] a ball of gold would surely not outstrip one of copper by as much as four fingers. Having observed this, I came to the conclusion that in a medium totally void of resistance all bodies would fall with the same speed [together]" [19]; thus Galileo had grasped an equivalence in gravity. The last conjecture is the famous Galileo equivalence principle; it serves as the beginning of our understanding of gravity. More precisely, Galileo equivalence principle states that in a gravitational field, the trajectory of a test body with a given initial velocity is independent of its internal structure and composition (universality of free fall trajectories).

From Galileo's observations, one can arrive at the following two well-known conclusions:

(a) The gravitational force (weight) at the top of the inclined plane and that at a middle point of the inclined plane can be regarded the same to the experimental limits in those days. Hence a falling body experiences a constant force (its weight). The motion of a falling body is one of constant acceleration. Therefore a constant force $f$ induces a motion of constant acceleration $a$. Hence force and acceleration (not velocity) are closely related. If one changes the inclinations of the plane to get different "dilutions" of gravity, one finds

$$f \propto a \qquad (18)$$

for a falling body. From Galileo's observation of the universality of free fall trajectories, we know that the acceleration $a$ is the same for different bodies. But $f$ (weight) is proportional to mass m. Hence for different bodies,

$$f/m \propto a. \tag{19}$$

If one chooses appropriate units, one arrives at

$$f = m\,a \tag{20}$$

for falling bodies. If one further assumes that all kind of forces are equivalent in their ability to accelerate and notices the vector nature of forces and accelerations, one would arrive at Newton's second law,

$$\boldsymbol{f} = m\,\boldsymbol{a}. \tag{21}$$

(b) From Galileo equivalence principle, the gravitational field can be described by the acceleration of gravity $\boldsymbol{g}$. Newton's second law for N particles in external gravitational field $\boldsymbol{g}$ is

$$m_I\, d^2\boldsymbol{x}_I/dt^2 = m_I\, \boldsymbol{g}(\boldsymbol{x}_I) + \sum_{J=1}^{N} \boldsymbol{F}_{IJ}(\boldsymbol{x}_I - \boldsymbol{x}_J), \qquad (I = 1,\ldots, N;\ J \neq I) \tag{22}$$

where $\boldsymbol{F}_{IJ}$ is the force acting on particle $I$ by Particle $J$. At a point $\boldsymbol{x}_0$, expand $\boldsymbol{g}(\boldsymbol{x}_I)$ as follows

$$\boldsymbol{g}(\boldsymbol{x}_I) = \boldsymbol{g}_0 + \underline{\boldsymbol{\lambda}} \cdot (\boldsymbol{x}_I - \boldsymbol{x}_0). \tag{23}$$

Choosing $\boldsymbol{x}_0$ as origin and applying the following non-Galilean space-time coordinate transformation

$$\boldsymbol{x}' = \boldsymbol{x} - (1/2)\, \boldsymbol{g}_0\, t^2,\ t' = t, \tag{24}$$

(22) is transformed to

$$m_I\, d^2\boldsymbol{x}'_I/dt'^2 = \sum_{J=1}^{N} \boldsymbol{F}_{IJ}(\boldsymbol{x}'_I - \boldsymbol{x}'_J) + O(\boldsymbol{x}'_K), \quad (I = 1,\ldots, N;\ J \neq I). \tag{25}$$

Thus we see that locally the effect of external gravitational field can be transformed away. Thus we arrive at a strong equivalence principle. Therefore in Newtonian mechanics,

Galileo Weak Equivalence Principle ⇔ Strong Equivalence Principle.

In the days of Galileo and Newton, the nature of light and radiation was controversial and had to wait for further development to clarify it.

(ii) *The second weak equivalence principle* (WEP II)

Since the motion of a macroscopic test body is determined not only by its trajectory but also by its rotation state, we have proposed from our previous studies [20, 21] the following stronger weak equivalence principle to be tested by experiments, which states that in a gravitational field, the motion of a test body with a given initial motion state is independent of its internal structure and composition (universality of free fall motions). By a test body, we mean a macroscopic body whose size is small compared to the length scale of the inhomogeneities of the gravitational field. The macroscopic body can have an intrinsic angular momentum (spin) including net quantum spin.

**2.3.** *Equivalence principles for photons (wave packets of light)*

(i) *WEP I for photons* (wave packets of light):

In analogue to the Galileo equivalence principle for test bodies, the WEP I for photons states that the spacetime trajectory of light in a gravitational field depends only on its initial position and direction of propagation, does not depend on its frequency (energy) and polarization.

(ii) *WEP II for photons* (wave packets of light):

The trajectory of light in a gravitational field depends only on its initial position and direction of propagation, not dependent of its frequency (energy) and polarization; the polarization state of the light will not change, e.g. no polarization rotation for linear polarized light; and no amplification/attenuation of light.

N.B. We consider the propagation (or trajectory) in eikonal approximation, i.e. in geometrical optics approximation. The wavelength must be small (just like a test body) than the inhomogeneity scale of the gravitational field.

**2.4.** *Microscopic equivalence principles*

The development of physics in the nineteenth century brought to improved understanding of light and radiations and to the development of special relativity. In 1905, Einstein [22] obtained the equivalence of mass and energy and derived the famous Einstein formula $E = mc^2$. A natural question came in at this point: How light and radiations behave in a gravitational field? In 1891, R. v. Eötvös [23] experiment showed that inertial mass and gravitational mass are equal to a high precision of $10^{-8}$. In June, 1907, Planck [24] reasoned that since all energy has inertial properties, all energies must gravitate. This paved the way to include the energy in the formulation of equivalence principle.

N.B. Since the power of EEP only reaches the gradient of gravity potential, it applies only to a region where the second-order gradients or curvature can be neglected. In applying the equivalence principle to wave packets or a microscopic wave function, we have to assume that the extension is limited to such a region. For example, it should not be applied to a long-distance entangled state.

(i) *Einstein equivalence principle* (EEP)

Two years after the proposal of special relativity and the formula $E=mc^2$, six months after Planck reasoned that all energy must gravitate, Einstein [25], in the last part (Principle of Relativity and Gravitation) of his comprehensive 1907 essay on relativity, proposed the complete physical equivalence of a homogeneous gravitational field to a uniformly accelerated reference system: "We consider two systems of motion, $\Sigma_1$ and $\Sigma_2$. Suppose $\Sigma_1$ is accelerated in the direction of its X axis, and $\gamma$ is the magnitude (constant in time) of this acceleration. Suppose $\Sigma_2$ is at rest, but situated in a homogeneous gravitational field, which imparts to all objects an acceleration $-\gamma$ in the direction of the X axis. As far as we know, the physical laws with respect to $\Sigma_1$ do not differ from those with respect to $\Sigma_2$, this derives from the fact that all bodies are accelerated alike in the gravitational field. We have therefore no reason to suppose in the present state of our experience that the systems $\Sigma_1$ and $\Sigma_2$ differ in any way, and will therefore assume in what follows the complete physical equivalence of the gravitational field and the corresponding acceleration of the reference system."[a] From this equivalence, Einstein derived clock and energy redshifts in a gravitational field. When applied to a spacetime region where inhomogeneities of the gravitational field can be neglected, this equivalence dictates the behavior of matter in gravitational field. The postulate of this equivalence is called the Einstein Equivalence Principle (EEP). EEP is the cornerstone of the gravitational coupling of matter and non-gravitational fields in general relativity and in metric theories of gravity.

EEP is a microscopic principle and may mean slightly different things for different people. To most people, EEP is equivalent to the coma-goes-to-semicolon rule for matter (not including gravitational energy) in gravitational field. Therefore, EEP means that in any and every local Lorentz (inertial) frame, anywhere and anytime in the universe, all the (nongravitational) laws of physics must take on their familiar special-relativistic forms [26]. *That is, local (nongravitational) physics should be universally special relativistic.* In other words, EEP says that the outcome of any local, nongravitational test experiment is independent of the velocity of the apparatus. For example, the fine structure constant $\alpha = e^2/\hbar c$ must be independent of location, time, and velocity.

___________

[a]Einstein further clarified the application of this equivalence to inhomogeneous field, e.g., in his book 'The Meaning of Relativity' (p. 58, Fifth edition, Princiton University Press 1955): '… We may look upon the principle of inertia as established, to a high degree of approximation, for the space of our planetary system, provided that we neglect the perturbations due to the sun and planets. Stated more exactly, there are finite regions, where, with respect to a suitably chosen space of reference, material particles move freely without acceleration, and in which the laws of special relativity, which have been developed above, hold with remarkable accuracy. Such regions we shall call "Galilean regions." We shall proceed from the consideration of such regions as a special case of known properties.'

(ii) *Modified Einstein equivalence principle* (MEEP)

In 1921, Eddington [27] mentioned the notion of an asymmetric affine connection in discussing possible extensions of general relativity. In 1922, Cartan [28] introduced torsion as the anti-symmetric part of an asymmetric affine connection and laid the foundation of this generalized geometry. Cartan [29] proposed that the torsion of spacetime might be connected with the intrinsic angular momentum of matter. In 1921-22, Stern and Gerlach [30] discovered the space quantization of atomic magnetic moments. In 1925-26, Goudsmit and Uhlenbeck [31] introduced our present concept of electron spin as the culmination of a series of studies of doublet and triplet structures in spectra. Following the idea of Cartan, Sciama [32, 33] and Kibble [34] developed a theory of gravitation which is commonly called the Einstein-Cartan-Sciama-Kibble (ECSK) theory of gravity.

After the works of Utiyama [35], Sciama [32, 33] and Kibble [34], interest and activities in gauge-type and torsion-type theories of gravity have continuously increased. Various different theories postulate somewhat different interaction of matter with gravitational field(s). In ECSK theory, in Poincaré gauge theories [36, 37] and in some other torsion theories, there is a torison gravitational field besides the usual metric field [38]. In special relativity, if we use a nonholonomic tetrad frame, there is an antisymmetrie part of the affine connection. Therefore many people working on torsion theory take the equivalence principle to mean something different from EEP so that torsion can be included. This is most clearly stated in P. von der Heyde's article "The Equivalence Principle in the $U_4$ Theory of Gravitation" [39]: *Locally the properties of special relativistic matter in a noninertial frame of reference cannot be distinguished from the properties of the same matter in a corresponding gravitational field*. This modified equivalence principle (MEEP) allows for formal inertial effects in a nonholonomic tetrad frame and hence allows torsion. There are two ways to treat the level of coupling of torsion; one can consider torsion on the same level as symmetric affine connection (MEEP I) or one can consider torsion on the same level as curvature tensors (MEEP II). Hehl, and von der Heyde [39] hold the second point of view. MEEP I allows torsion. Since torsion is a tensor, it cannot be transformed away in any frame if it is not zero. EEP is equivalent to MEEP I plus no torsion; therefore we have EEP implies MEEP I but MEEP I does not implies EEP. For a test body, curvature effects are neglected; so MEEP II is essentially equivalent to EEP for test bodies. Test bodies with nonvanishing total intrinsic spin feel torques from the torsion field. Hence MEEP I does not imply WEP II. Moreover MEEP I does not imply WEP I either [40]. Therefore we have the following:

$$\begin{array}{ccc} \text{EEP} & \Rightarrow & \text{MEEP I} \\ *\Uparrow\Downarrow & \rightarrowtail & \leftarrowtail \\ \text{WEP II} & \Rightarrow & \text{WEP I} \end{array}$$

*WEP II implies EEP is proved for an electromagnetic system in $\chi$-$g$ framework [20, 21, 41]. However, for other frameworks, the issue is still open.

## 2.5. *Equivalence principles including gravity (Strong equivalence principles)*

How does gravitational energy behave in a gravitational field? Is local gravity experiment depending on where and when in the universe it is performed? These involve nonlinear gravity effects.

(i) *WEP I for massive bodies*

This weak equivalence principle says that in a gravitational field, the trajectory of a massive test body with a given initial velocity is also independent of the amount of gravitational self-energy inside the massive body. In Brans-Dicke theory and many other theories, there are violations of this equivalence principle. The violations are called Nordtvedt effects [42, 43]. General relativity obeys WEP I for massive bodies in the post-Newtonian limit and for black hole solutions. The nonexistence of Nordtvedt effects is an efficient way to single out purely metric theory among metric theories of gravity (those comply with EEP). From lunar laser ranging experiment and binary pulsar timing observations, the Nordtvedt effect is limited.

(ii) Dicke's [44] strong equivalence principle (SEP)

This is a microscopic equivalence principle. It says that the outcome of any local test experiment -- gravitational or nongravitational -- is independent of where and when in the universe it is performed, and independent of the velocity of the apparatus. If this equivalence principle is valid, the Newtonian gravitational constant $G_N$ should be a true constant. Brans-Dicke theory with its variable "gravitational constant" as measured by Cavendish experiments satisfies EEP but violates SEP. Also, if this equivalence principle is valid, a self-gravitating system in background with length scale much larger than the self-gravitating system should have locally Lorentz invariance in the background, e.g., no preferred-frame effects [45, 46].

The violations of SEP seem to be linked with the violations of WEP I for massive bodies in many cases. It is interesting to know how SEP and WEP I for massive bodies are connected. The violations of SEP may also be connected to the violations of WEP I at some level in some cases.

We note in passing that there are other versions of equivalence principles which we are not able to list them here one-by-one. For recent discussions on equivalence principles, see also [47, 48].

## 2.4. *Inequivalence and interrelations of various equivalence principles*

In the preceding subsections, we have listed and explained various equivalence principles. Logically all these equivalence principles are different. An important issue is that to what extent they are equivalent, and in what situations they are inequivalent. This issue became conspicuous for more than 50 years since Dicke-Schiff redshift controversy. In 1960, Leonard Schiff [49] argued as follows: "The Eötvös experiments show with considerable accuracy that the gravitational and inertial masses of normal matter are equal. This means that the ground state eigenvalue of the Hamiltonian for this matter appears equally in the inertial mass and in the interaction of this mass with a gravitational field. It would be quite remarkable if this could occur without the entire Hamiltonian being involved in the same way, in which case a clock composed of atoms whose motions are determined by this Hamiltonian would have its rate affected in the expected manner by a gravitational field." He suggested that EEP and, hence, the metric gravitational redshift are consequences of WEP I. In short, Schiff believes that

$$\text{WEP I} \Rightarrow \text{EEP}.$$

This conjecture is known as Schiff's conjecture. The scope of validity of Schiff's conjecture has great importance in the analysis of the empirical foundations of EEP.

However Dicke [50] held a different point of view and believed that the redshift experiment has independent theoretical significance. In November 1970, the interests in the issue of the validity of Schiff's conjecture were rekindled during a vigorous argument between L. Schiff and K.S. Thorne at the Caltech-JPL Conference on Experimental Tests of Gravitation Theories. In 1973, Thorne, Lee, and Lightman [51] analyzed the fundamental concepts and terms involved in detail and gave a plausibility argument supporting Schiff's conjecture. Lightman and Lee [52] proved Schiff's conjecture for electromagnetically interacting systems in a static, spherically symmetric gravitational field using the $TH\varepsilon\mu$ formalism. I found a nonmetric theory which includes pseudoscalar-photon interaction and showed that it is a counterexample to Schiff's conjecture [53]. In 1974, I showed that this counterexample is the only case in a general premetric constitutive tensor formulation of electromagnetism ($\chi$-framework) with standard particle Lagrangian (The whole framework is called the $\chi$-$g$ framework.) [20, 21]. This supports that the approach of Schiff is right in the large, although not completely right. In the eikonal approximations of the $\chi$-$g$ framework, I showed that the first-order gravitational redshifts are metric [21] (so Schiff was right for redshift in this case to first order). In the latter part of 1970's, I use the $\chi$-g framework to look into the issue of gravitational coupling to electromagnetism empirically [14-16, 40]. In the next section, we will review the progress for this issue. Recently, the significance of redshift experiments is brought up again in the comparison of redshift and atom interferometry experiment [54, 55].

For the strong equivalence principle, one could ask similar questions. Would WEP I for massive body imply Dicke's strong equivalence principle (SEP)? This is a direct extension of Schiff's conjecture. One can call it Schiff's conjecture for massive bodies. There are significant progresses recently. Gérard [56] has worked out a link between the vanishing of Nordtvedt effects and a condition of SEP. Di Casola, Leberati and Sonego [57] have employed WEP I for massive bodies as a sieve for purely metric theories of gravity using variational approach. They also propose the conjecture that SEP is equivalent to the union of WEP I for massive bodies (GWEP in their term) and EEP. Since WEP I does not imply EEP (Schiff's conjecture is incorrect) [20, 21, 53], we would like to propose to investigate the validity of the following two statements in various frameworks: (i) WEP II for massive bodies is equivalent to Dicke's strong equivalence principle (SEP); (ii) SEP is equivalent to the union of WEP II for massive bodies (GWEP in their term) and EEP.

## 3. Gravitational Coupling to Electromagnetism and the Structure of Spacetime

### 3.1. Premetric electrodynamics as a framework to study gravitational coupling to electromagnetism

For the ordinary gravitational field, it is a low energy situation compared to Planck energy, as we mentioned in the Introduction. If we represent the gravitational coupling to electromagnetism by constitutive tensor density, the constitutive tensor density must be linear and local as given by (12), independent of the field strength $F_{kl}$, dependent only on the gravitational field(s). The constitutive tensor density (12) has three irreducible pieces. Both $H^{ij}$ and $F_{kl}$ are antisymmetric, hence $\chi^{ijkl}$ must be antisymmetric in $i$ and $j$, and $k$ and $l$. Therefore the constitutive tensor density $\chi^{ijkl}$ has 36 (6 × 6) independent components. A general linear constitutive tensor density $\chi^{ijkl}$ in electrodynamics can first be decomposed into two parts, the symmetric part in the exchange of index pairs $ij$ and $kl$ [(1/2) ($\chi^{ijkl}$ + $\chi^{klij}$ )] and the antisymmetric part in the exchange of index pairs $ij$ and $kl$ [(1/2) ($\chi^{ijkl}$ − $\chi^{klij}$ )]. The first part has 21 degrees of freedom and contains the totally antisymmetric part – the axion part (Ax). Subtracting the axion part, the remaining part is the principal part which has 20 degrees of freedom. The second part is the skewon part and has 15 degrees of freedom. The principal part (P), the axion part (Ax) and the Hehl-Obukhov-Rubilar skewon part (Sk) constitute the three irreducible parts under the group of general coordinate transformations [6]:

$$\chi^{ijkl} = {}^{(P)}\chi^{ijkl} + {}^{(Sk)}\chi^{ijkl} + {}^{(Ax)}\chi^{ijkl}, \quad (\chi^{ijkl} = -\chi^{jikl} = -\chi^{ijlk}) \qquad (26)$$

with

$$\begin{align}
{}^{(P)}\chi^{ijkl} &= (1/6)[2(\chi^{ijkl} + \chi^{klij}) - (\chi^{iklj} + \chi^{ljik}) - (\chi^{iljk} + \chi^{jkil})], & (27a)\\
{}^{(Ax)}\chi^{ijkl} &= \chi^{[ijkl]} = \varphi\, e^{ijkl}, & (27b)\\
{}^{(Sk)}\chi^{ijkl} &= (1/2)\,(\chi^{ijkl} - \chi^{klij}). & (27c)
\end{align}$$

Decomposition (26) is unique. If we substitute (26) into (17a), the skewon part does not contribute to the Lagrangian; hence, for Lagrangian based theory, it is skewonless. The systematic study of skewonful cases started in 2002 (See, e.g., Hehl and Obukhov [6]).

The complete agreement with EEP for photon sector requires (as locally in special relativity) (i) no birefringence; (ii) no polarization rotation; (iii) no amplification/no attenuation in spacetime propagation. In sections 3.2 to 3.5 we review how cosmic connection/observation of these three conditions on electromagnetic propagation verifies EEP and determination of the spacetime structure in the skewonless case (Lagrangian-based case). In section 3.2, we derive wave propagation and dispersion relations in the lowest eikonal approximation in weak field in the premetric electrodynamics. In section 3.3, we apply it to the determination of the spacetime structure in the skewonless case using no birefringence condition. With no birefringence, any skewonless spacetime constitutive tensor must be of the form

$$\chi^{ijkl} = (-h)^{1/2}[(1/2)h^{ik} h^{jl} - (1/2)h^{il} h^{kj}]\psi + \varphi e^{ijkl}, \tag{28}$$

where $h^{ij}$ is a metric constructed from $\chi^{ijkl}$ ($h = \det(h_{ij})$ and $h_{ij}$ the inverse of $h^{ij}$) which generates the light cone for electromagnetic wave propagation, $\psi$ a dilaton field constructed from $\chi^{ijkl}$ and $\varphi$ an axion field constructed from $\chi^{ijkl}$. Observations on no birefringence of cosmic propagation of electromagnetic waves constrain the spacetime constitutive tensor to the form (28) to very high precision. In section 3.4, we review the derivation of the dispersion relation of wave propagation in dilaton field and axion field with constitutive relation (28); we show further that with the condition of no polarization rotation and the condition of no amplification/no attenuation satisfied, the axion $\varphi$ and the dilaton $\psi$ should be constant, i.e. no varying axion field and no varying dilaton field respectively. The EEP for photon sector would then be observed; the spacetime constitutive tensor density would be of metric-induced form. Thus we tie the three observational conditions to EEP and to metric-induced spacetime constitutive tensor density in the photon sector. In section 3.5, we review the empirical constraint on cosmic dilaton field and cosmic axion field. The results are summarized in Table I in the Introduction. In section 3.6, we apply the dispersion relations derived in section 3.2 to the case of metric induced constitutive tensor with skewons with further discussions. In section 3.7, we discuss the case of spacetime with asymmetric-metric induced constitutive tensor using Fresnel equaion. In section 3.8, we review the application of these results to the accuracy of empirical verification of the closure relations in electrodynamics.

### 3.2. Wave propagation and the dispersion relation

The sourceless Maxwell equation (10b) is equivalent to the local existence of a 4-potential $A_i$ such that

$$F_{ij} = A_{j,i} - A_{i,j}, \tag{29}$$

with a gauge transformation freedom of adding an arbitrary gradient of a scalar function to $A_i$. The Maxwell equation (10a) in vacuum is

$$(\chi^{ijkl}A_{k,l})_{,j} = 0. \tag{30}$$

Using the derivation rule, we have

$$\chi^{ijkl}A_{k,lj} + \chi^{ijkl}{}_{,j} A_{k,l} = 0. \tag{31}$$

(i) For slowly varying, nearly homogeneous field/medium, and/or (ii) in the eikonal

approximation with typical wavelength much smaller than the gradient scale and time-variation scale of the field/medium, the second term in (31) can be neglected compared to the first term, and we have

$$\chi^{ijkl} A_{k,lj} = 0. \tag{32}$$

This approximation is the lowest eikonal approximation, usually also called the eikonal approximation. In this approximation, the dispersion relation is given by the generalized covariant quartic Fresnel equation (see, e.g. [6]; also section 3.7). It is well-known that axion does not contribute to this dispersion relation [6, 14-16, 58-61] as we will see in the following. In this subsection, we use this lowest eikonal approximation and follow Ref. [62] to derive dispersion relation in the general linear local constitutive framework. In the subsection 3.4, we keep the second term of (31) and follow Ref. [63] to find out dispersion relations for the case that the dilaton gradient and the axion gradient cannot be neglected.

In the weak field or dilute medium, we assume

$$\chi^{ijkl} = \chi^{(0)ijkl} + \chi^{(1)ijkl} + O(2), \tag{33}$$

where O(2) means second order in $\chi^{(1)}$. Since the violation from the Einstein Equivalence Principle would be small and/or if the medium is dilute, in the following we assume that

$$\chi^{(0)ijkl} = (1/2) g^{ik} g^{jl} - (1/2) g^{il} g^{kj}, \tag{34}$$

and $\chi^{(1)ijkl}$ is small compared with $\chi^{(0)ijkl}$. We can then find a locally inertial frame such that $g^{ij}$ becomes the Minkowski metric $\eta^{ij}$ good to the derivative of the metric. To look for wave solutions, we use eikonal approximation and choose $z$-axis in the wave propagation direction so that the solution takes the following form:

$$A = (A_0, A_1, A_2, A_3)\, e^{ikz - i\omega t}. \tag{35}$$

We expand the solution as

$$A_i = [A^{(0)}{}_i + A^{(1)}{}_i + O(2)]\, e^{ikz - i\omega t}. \tag{36}$$

Imposing radiation gauge condition in the zeroth order in the weak field/dilute medium/weak EEP violation approximation, we find the zeroth order solution of (36) and the zeroth order dispersion relation satisfying the zeroth order equation $\chi^{(0)ijkl} A^{(0)}{}_{k,lj} = 0$ as follow:

$$A^{(0)} = (0, A^{(0)}{}_1, A^{(0)}{}_2, 0), \quad \omega = k + O(1). \tag{37}$$

Substituting (36) and (37) into equation (32), we have

$$\chi^{(1)ijkl} A^{(0)}{}_{k,lj} + \chi^{(0)ijkl} A^{(1)}{}_{k,lj} = 0 + O(2). \tag{38}$$

The $i = 0$ and $i = 3$ components of (38) both give

$$A^{(1)}{}_0 + A^{(1)}{}_3 = 2\,(\chi^{(1)3013} - \chi^{(1)3010})\, A^{(0)}{}_1 + 2\,(\chi^{(1)3023} - \chi^{(1)3020})\, A^{(0)}{}_2 + O(2). \tag{39}$$

Since this equation does not contain $\omega$ and $k$, it does not contribute to the determination of the dispersion relation. A gauge condition in the O(1) order fixes the values of $A^{(1)}{}_0$ and $A^{(1)}{}_3$.

The $i = 1$ and $i = 2$ components of (38) are

$$(1/2)(\omega^2 - k^2)\, A^{(0)}{}_1 + \chi^{(0)1jkl} A^{(1)}{}_{k,lj} + \chi^{(0)1jkl} A^{(0)}{}_{k,lj} = 0 + O(2), \tag{40a}$$
$$(-1/2)(\omega^2 - k^2)\, A^{(0)}{}_2 + \chi^{(0)2jkl} A^{(1)}{}_{k,lj} + \chi^{(1)2jkl} A^{(0)}{}_{k,lj} = 0 + O(2). \tag{40b}$$

These two equations determine the dispersion relation and can be rewritten as

$$[(1/2)(\omega^2 - k^2) - k^2 A_{(1)}]\, A^{(0)}{}_1 - k^2 B_{(1)}\, A^{(0)}{}_2 = O(2), \tag{41a}$$
$$-k^2 B_{(2)}\, A^{(0)}{}_1 + [(1/2)(\omega^2 - k^2) - k^2 A_{(2)}]\, A^{(0)}{}_2 = O(2), \tag{41b}$$

where

$$A_{(1)} \equiv \chi^{(1)1010} - (\chi^{(1)1013} + \chi^{(1)1310}) + \chi^{(1)1313}, \tag{42a}$$
$$A_{(2)} \equiv \chi^{(1)2020} - (\chi^{(1)2023} + \chi^{(1)2320}) + \chi^{(1)2323}, \tag{42b}$$
$$B_{(1)} \equiv \chi^{(1)1020} - (\chi^{(1)1023} + \chi^{(1)1320}) + \chi^{(1)1323}, \tag{42c}$$
$$B_{(2)} \equiv \chi^{(1)2010} - (\chi^{(1)2013} + \chi^{(1)2310}) + \chi^{(1)2313}. \tag{42d}$$

We note that *$A_{(1)}$ and $A_{(2)}$ contain only the principal part of $\chi$; $B_{(1)}$ and $B_{(2)}$ contain only the principal and skewon part of $\chi$. The axion part drops out and does not contribute to the dispersion relation in the eikonal approximation.* The principal part $^{(P)}B$ and skewon part $^{(Sk)}B$ of $B_{(1)}$ are as follows:

$$^{(P)}B = (1/2)(B_{(1)} + B_{(2)}); \quad ^{(Sk)}B = (1/2)(B_{(1)} - B_{(2)}). \tag{43}$$

From (43), $B_{(1)}$ and $B_{(2)}$ can be expressed as

$$B_{(1)} = {}^{(P)}B + {}^{(Sk)}B; \quad B_{(2)} = {}^{(P)}B - {}^{(Sk)}B. \tag{44}$$

For equations (41a,b) to have nontrivial solutions of $(A_1^{(0)}, A_2^{(0)})$, we must have the following determinant vanish to first order:

$$\det \begin{bmatrix} (1/2)(\omega^2 - k^2) - k^2 A_{(1)}] & -k^2 B_{(1)} \\ -k^2 B_{(2)} & (1/2)(\omega^2 - k^2) - k^2 A_{(2)} \end{bmatrix}$$
$$= (1/4)(\omega^2 - k^2)^2 - (1/2)(\omega^2 - k^2)\, k^2 (A_{(1)} + A_{(2)}) + k^4\, (A_{(1)} A_{(2)} - B_{(1)} B_{(2)}) = 0 + O(2). \tag{45}$$

The solution of this quadratic equation in $\omega^2$, i.e., the dispersion relation is

$$\omega^2 = k^2[1 + (A_{(1)} + A_{(2)}) \pm ((A_{(1)} - A_{(2)})^2 + 4B_{(1)} B_{(2)})^{1/2}] + O(2), \qquad (46)$$

or

$$\omega = k [1 + 1/2 (A_{(1)} + A_{(2)}) \pm 1/2 ((A_{(1)} - A_{(2)})^2 + 4B_{(1)} B_{(2)})^{1/2}] + O(2). \qquad (47)$$

From (46) the group velocity is

$$v_g = \partial\omega/\partial k = 1 + 1/2 (A_{(1)} + A_{(2)}) \pm 1/2 ((A_{(1)} - A_{(2)})^2 + 4B_{(1)} B_{(2)})^{1/2} + O(2). \qquad (48)$$

The quantity under the square root sign is

$$\xi \equiv (A_{(1)} - A_{(2)})^2 + 4B_{(1)} B_{(2)} = (A_{(1)} - A_{(2)})^2 + 4(^{(P)}B)^2 - 4(^{(Sk)}B)^2. \qquad (49)$$

Depending on the sign or vanishing of $\xi$, we have the following three cases of electromagnetic wave propagation:

(i) $\xi > 0$, $(A_{(1)} - A_{(2)})^2 + 4(^{(P)}B)^2 > 4(^{(Sk)}B)^2$: There is birefringence of wave propagation;
(ii) $\xi = 0$, $(A_{(1)} - A_{(2)})^2 + 4(^{(P)}B)^2 = 4(^{(Sk)}B)^2$: There are no birefringence and no dissipation/amplification in wave propagation;
(iii) $\xi < 0$, $(A_{(1)} - A_{(2)})^2 + 4(^{(P)}B)^2 < 4(^{(Sk)}B)^2$: There is no birefringence, but there are both dissipative and amplifying modes in wave propagation.

### 3.2.1. The condition of vanishing of $B_{(1)}$ and $B_{(2)}$ for all directions of wave propagation

From the definition (29c), the condition of vanishing of $B_{(1)}$ for wave propagation in the $z$-axis direction is

$$B_{(1)} = \chi^{(1)1020} + \chi^{(1)1323} - \chi^{(1)1023} - \chi^{(1)1320} = 0. \qquad (50)$$

To look for conditions derivable in combination with those from other directions, we do active Lorentz transformations (rotations/boosts). Active rotation $R_\theta$ in the $y$-$z$ plane with angle $\theta$ is

$$\underline{t} = R_\theta t,\ \underline{x} = R_\theta x,\ \underline{y} = R_\theta y = y \cos\theta + z \sin\theta,\ \underline{z} = R_\theta z = - y \sin\theta + z \cos\theta. \qquad (51)$$

Applying active rotation $R_\theta$ (51) to (50), we have

$$\begin{aligned}0 &= \chi^{(1)1020} + \chi^{(1)1323} - \chi^{(1)1023} - \chi^{(1)1320} \\ &= \chi^{(1)1020} + \chi^{(1)1323} - \chi^{(1)1023} - \chi^{(1)1320} + \theta\,(\chi^{(1)1030} + \chi^{(1)1220} - \chi^{(1)1223} - \chi^{(1)1330}) + O(\theta^2),\end{aligned} \qquad (52)$$

for small value of $\theta$. From (52) and (50), we have

$$\chi^{(1)1030} + \chi^{(1)1220} - \chi^{(1)1223} - \chi^{(1)1330} = 0. \tag{53}$$

Following the same procedure, we apply repeatedly active rotation $R_\theta$ to (53) and the resulting equations together with their linear combinations. After performing cyclic permutation 1→2→3→1 on the upper indices once and twice on some of the resulting equations, we have the following equations (for detailed derivation, see arXiv:1312.3056v1)

$$\chi^{(1)1220} = \chi^{(1)1330}; \chi^{(1)2330} = \chi^{(1)2110}; \chi^{(1)3110} = \chi^{(1)3220}; \chi^{(1)1020} = -\chi^{(1)1323}; \chi^{(1)2030} = -\chi^{(1)2131};$$
$$\chi^{(1)3010} = -\chi^{(1)3212}; \chi^{(1)1320} = -\chi^{(1)1230}; \chi^{(1)3210} = -\chi^{(1)3120}; \chi^{(1)2130} = -\chi^{(1)2310};$$
$$\chi^{(1)1023} = -\chi^{(1)1320}; \chi^{(1)2031} = -\chi^{(1)2130}; \chi^{(1)3012} = -\chi^{(1)3210}. \tag{54a-l}$$

From (54g-l), $\chi^{(1)0123}$ is completely anti-symmetric under any permutation of (0123). Among (54g-i) only 2 are independent; among (54j-l) also only 2 are independent. For $^{(PA)}\chi^{ijkl}$, (54g-l) give 2 independent conditions. For $^{(Sk)}\chi^{ijkl}$, (54g-l) give 3 independent conditions and $\chi^{(1)0123}$ must vanish.

The derivation of formulas in this subsection from (50) to (54l) is independent of whether $\chi^{ijkl}$ is principal, axionic or skewonic. Hence, $^{(P)}$(54a-l) hold for $^{(P)}\chi^{ijkl}$ with $^{(P)}B_{(1)}$ = 0, $^{(A)}$(54a-l) hold for $^{(A)}\chi^{ijkl}$ with $^{(A)}B_{(1)}$ = 0, and $^{(Sk)}$(54a-l) holds for $^{(Sk)}\chi^{ijkl}$ with $^{(Sk)}B_{(1)}$ = 0. Here $^{(P)}$(54a-l) means (54a-l) with $\chi$ substituted by $^{(P)}\chi$, $^{(A)}$(54a-l) means (54a-l) with $\chi$ substituted by $^{(A)}\chi$, and $^{(Sk)}$(54a-l) means (54a-l) with $\chi$ substituted by $^{(Sk)}\chi$; similarly for $^{(P)}B_{(1)}$, $^{(A)}B_{(1)}$ and $^{(Sk)}B_{(1)}$. For $B_{(1)} = B_{(2)} = 0$ in all directions, we have $^{(P)}B_{(1)} = {}^{(Sk)}B_{(1)} = 0$ in all directions, and hence, both $^{(P)}$(54a-l) and $^{(Sk)}$(54a-l) are valid.

*3.2.2. The condition of $^{(Sk)}B_{(1)} = {}^{(P)}B_{(1)} = 0$ and $A_{(1)} = A_{(2)}$ for all directions of wave propagation*

With the condition $^{(Sk)}B_{(1)} = {}^{(P)}B_{(1)} = 0$ and $A_{(1)} = A_{(2)}$ for all directions of wave propagation, there is no birefringence for all directions of wave propagation. From subsection 2.2.1, we have equations (54a-l) holds from the validity of $^{(Sk)}B_{(1)} = {}^{(P)}B_{(1)} = 0$ (i.e., $B_{(1)} = 0$) for all directions of wave propagation. From $A_{(1)} = A_{(2)}$ and the definition (42a, b), we have

$$\chi^{(1)1010} - (\chi^{(1)1013} + \chi^{(1)1310}) + \chi^{(1)1313} = \chi^{(1)2020} - (\chi^{(1)2023} + \chi^{(1)2320}) + \chi^{(1)2323}. \tag{55}$$

From (54c) for the principal part, the terms in the parentheses on the two sides of the above equation cancel out and we have

$$\chi^{(1)1010} + \chi^{(1)1313} = \chi^{(1)2020} + \chi^{(1)2323}. \tag{56a}$$

Applying active rotation $R_{\pi/2}$ around in the y-z plane to (56a), we obtain

$$\chi^{(1)1010} + \chi^{(1)1212} = \chi^{(1)3030} + \chi^{(1)3232}. \tag{56b}$$

### 3.3. Nonbirefringence condition for the skewonless case

If EEP is observed, photons with different polarizations as test particles shall follow identical trajectories in a gravitational field. Then the photons obey WEP I and there is no birefringence. In this section, we will first derive the core metric formula for the constitutive tensor density from the nonbirefringence condition in the skewonless case (Lagrangian-based case) and then use the cosmological observations to constrain the spacetime constitutive tensor density to this form to ultra-high precision.

From equation (49) in section 3.2, the condition of nonbirefringence in the skewonless case is

$$A_{(1)} = A_{(2)}, B_{(1)} = B_{(2)} = {}^{(P)}B = 0. \tag{57}$$

With these conditions, (54a-h) and (56a,b) in section 2.2.1 and section 2.2.2 are valid and gives ten conditions on 21 independent components of skewonless constitutive tensor density $\chi^{ijkl}$:

$$\chi^{(1)1220} = \chi^{(1)1330}; \tag{58a}$$
$$\chi^{(1)2330} = \chi^{(1)2110}; \tag{58b}$$
$$\chi^{(1)3110} = \chi^{(1)3220}; \tag{58c}$$
$$\chi^{(1)1020} = -\chi^{(1)1323}; \tag{58d}$$
$$\chi^{(1)2030} = -\chi^{(1)2131}; \tag{58e}$$
$$\chi^{(1)3010} = -\chi^{(1)3212}; \tag{58f}$$
$$\chi^{(1)1320} = -\chi^{(1)1230}; \tag{58g}$$
$$\chi^{(1)3210} = -\chi^{(1)3120}; \tag{58h}$$
$$\chi^{(1)1010} + \chi^{(1)1313} = \chi^{(1)2020} + \chi^{(1)2323}; \tag{58i}$$
$$\chi^{(1)1010} + \chi^{(1)1212} = \chi^{(1)3030} + \chi^{(1)3232}. \tag{58j}$$

Define

$$h^{(1)10} \equiv h^{(1)01} \equiv -2\,{}^{(P)}\chi^{(1)1220};\ h^{(1)20} \equiv h^{(1)02} \equiv -2\,{}^{(P)}\chi^{(1)2330};\ h^{(1)30} \equiv h^{(1)03} \equiv -2\,{}^{(P)}\chi^{(1)3110};$$
$$h^{(1)12} \equiv h^{(1)21} \equiv -2\,{}^{(P)}\chi^{(1)1020};\ h^{(1)23} \equiv h^{(1)32} \equiv -2\,{}^{(P)}\chi^{(1)2030};\ h^{(1)31} \equiv h^{(1)13} \equiv -2\,{}^{(P)}\chi^{(1)3010};$$
$$h^{(1)11} \equiv 2\,{}^{(P)}\chi^{(1)2020} + 2\,{}^{(P)}\chi^{(1)2121} - h^{(1)00};\ h^{(1)22} \equiv 2\,{}^{(P)}\chi^{(1)3030} + 2\,{}^{(P)}\chi^{(1)3232} - h^{(1)00};$$
$$h^{(1)33} \equiv 2\,{}^{(P)}\chi^{(1)1010} + 2\,{}^{(P)}\chi^{(1)1313} - h^{(1)00}, \tag{59a}$$

$$\psi \equiv 1 + 2\,{}^{(P)}\chi^{(1)1212} + (1/2)\,\eta_{00}\,(h^{(1)00} - h^{(1)11} - h^{(1)22} - h^{(1)33}) - h^{(1)11} - h^{(1)22}, \tag{59b}$$

$$\varphi \equiv \chi^{(1)0123} \equiv \chi^{(1)[0123]}. \tag{59c}$$

Note that in these definitions, $h^{(1)00}$ is not defined and is free. Now it is straightforward to show that when (41a-l) and (46a-b) are satisfied, then $\chi$ can be written to first-order in terms of the fields $h^{(1)ij}$, $\psi$, $\varphi$, and $p^{ij}$ with $h^{ij} \equiv \eta^{ij} + h^{(1)ij}$ and $h \equiv \det(h_{ij})$ in the following

form:

$$\chi^{ijkl} = {}^{(P)}\chi^{(1)ijkl} + {}^{(A)}\chi^{(1)ijkl} + {}^{(SkII)}\chi^{(1)ijkl}$$
$$= \tfrac{1}{2}\,(-h)^{1/2}[h^{ik}h^{jl} - h^{il}h^{kj}]\psi + \varphi e^{ijkl}, \qquad (60)$$

with

$$^{(P)}\chi^{(1)ijkl} = \tfrac{1}{2}\,(-h)^{1/2}[h^{ik}h^{jl} - h^{il}h^{kj}]\psi, \qquad (61a)$$
$$^{(A)}\chi^{(1)ijkl} = \varphi e^{ijkl}. \qquad (61b)$$

It is ready to derive the following theorem:

*Theorem*: For linear electrodynamics with Lagrangian (17a), i.e. with skewonless constitutive relation (12), the following three statements are equivalent to first order in the field:
(i) $A_{(1)} = A_{(2)}$ and $^{(P)}B = 0$ for all directions, i.e. nonbirefringence in electromagnetic wave propagation,
(ii) (58a-j) hold,
(iii) $\chi^{ijkl}$ can be expressed as (60) with (59a-c).

*Proof*: (i) → (ii) has been demonstrated in the derivation of (58a-j).
 (ii) → (iii) has also been demonstrated in the derivation of (60) above.
 (iii) → (i): (60) is a Lorentz tensor density equation. If it holds in one Lorentz frame, it holds in any other frame. From this we readily check that $A_{(1)} = A_{(2)}$ and $^{(P)}B = 0$ in any new frame with the wave propagation in the *z*-direction.

This theorem is a re-statement of the results of our work [14-16]. We note that previously we used the symbol $H^{ik}$ instead of $h^{ik}$, here because $H^{ik}$ is already used for excitation, we changed the notation.

We constructed the relation (60) in the weak-violation approximation of EEP in 1981 [14-16]; Haugan and Kauffmann [58] reconstructed the relation (60) in 1995. After the cornerstone work of Lämmerzahl and Hehl [59], Favaro and Bergamin [64] finally proved the relation (60) without assuming weak-field approximation (see also Dahl [65]).

Polarization measurements of electromagnetic waves from pulsars and cosmologically distant astrophysical sources has yielded stringent constraints agreeing with (60) down to $10^{-16}$ and $10^{-32}$ respectively as shown in Table 1.

*Observational constraints from pulsars* [15, 16]: In 1970s and 1980s pulsar observations gave the best constraints on the birefringence in the propagation. The pulses and micropulses from pulsars with different polarizations are correlated in general structure and timing [66]. No retardation with respect to different polarizations is observed. This means that conditions similar to (57) are satisfied to observational accuracy. For Crab pulsar, the micropulses with different polarizations are correlated in timing to within $10^{-4}$ sec, the distance of the Crab pulsar is 2200 pc, therefore to within $10^{-4}$ sec / (2200 × 3.26 light yr.) = $5 \times 10^{-16}$ accuracy two conditions similar to (57) are satisfied. In 1981, over 300 pulsars in different directions had been observed. Many of them had polarization data. Combining all of them, (58a-j) were satisfied to an accuracy of $10^{-14} – 10^{-16}$. Since for galactic gravitational field $U \sim 10^{-6}$, according to the procedure of proving the theorem, $\chi^{(1)}/U$ (or $\chi/U$) agrees with that given by (60) to an accuracy of $10^{-8} – 10^{-10}$. At that time, we anticipated that detailed analysis would reveal better results. In 2002, a detailed analysis using X-ray pulsars [67] demonstrated the full procedure. At that time McCulloch, Hamilton, Ables and Hunt [68] had just observed a radio pulsar in the large Magellanic Cloud; Backer, Kulkarni, Helles, Davis and Goss [69] had discovered a millisecond pulsar which rotates 20 times faster than the Crab pulsar. The progress of these observations would potentially give better constraints on some of the conditions (58a-j) due to larger distance or fast period involved.

We also anticipated that analysis of optical and X-ray polarization data from various astrophysical sources would give better accuracy to some of the ten constraints in (58a-j).

Thus, to high accuracy, photons are propagating in the metric field $h^{ik}$ and two additional (pseudo)scalar fields $\psi$ and $\varphi$. A change of $h^{ik}$ to $\lambda h^{ik}$ does not affect $\chi^{ijkl}$ in (60) -- this corresponds to the freedom of $h^{(1)00}$ in the definition (59a) of $h^{(1)ij}$. Thus we have constrained the general linear constitutive tensor of 21 degrees of freedom from the 10 constraints (58a-j) to eleven degrees of freedom in (60).

*Constraints from extragalactic radio-galaxy observations* [61]: Analyzing the data from polarization measurements of extragalactic radio sources, Haugan and Kauffmann [58] in 1995 inferred that the resolution for null-birefringence is 0.02 cycle at 5 GHz. This corresponds to a time resolution of $4 \times 10^{-12}$ s and gives much better constraints. With a detailed analysis and more extragalactic radio observations, (60) would be tested down to $10^{-28}$-$10^{-29}$ at cosmological distances. In 2002, Kostelecky and Mews [70] used polarization measurements of light from cosmologically distant astrophysical sources to yield stringent constraints down to $2 \times 10^{-32}$. The electromagnetic propagation in Moffat's nonsymmetric gravitational theory [71, 72] fits the $\chi$-$g$ framework. Krisher [73], and Haugan and Kauffmann [58] have used the pulsar data and extragalactic radio observations respectively to constrain it.

*Constraints from gamma-ray burst observations* [17]: Recent polarization observations on gamma-ray bursts gives even better constraints on the dispersion relation and nonbirefringence in cosmic propagation [74, 75]. The observation on the polarized gamma-ray burst GRB 061122 ($z = 1.33$) gives a lower limit on its polarization fraction of 60% at 68% confidence level (c.l.) and 33% at 90% c.l. in the 250-800 keV energy range [27]. The observation on the polarized gamma-ray burst GRB 140206A constrains the linear polarization level of the second peak of this GRB above 28 % at 90% c.l. in the 200-400 keV energy range [75]; the redshift of the source is measured from the GRB afterglow optical spectroscopy to be $z = 2.739$. GRBs polarization observations have been used to set constraints on various dispersion relations (See, e.g. [76, 77] and references therein). These two new GRB observations have larger and better redshift determinations than previous observations. We use them to give better constraints in our case. Since birefringence is proportional to the wave vector $k$ in our case, as gamma-ray of a particular frequency (energy) travels in the cosmic spacetime, the two linear polarization eigen-modes would pick up small phase differences. A linear polarization mode from distant source resolved into these two modes will become elliptical polarized during travel and lose part of the linear coherence. The way of gamma ray losing linear coherence depends on the frequency span. For a band of frequency, the extent of losing coherence depends on the distance of travel. The depolarization distance is of the order of frequency band span $\pi\Delta f$ times the integral $I = \int(1 + z(t))dt$ of the redshift factor $(1 + z(t))$ with respect to the time of travel. For GRB 140206A, this is about

$$\pi\Delta f\, I = \pi\Delta f \int(1 + z(t))dt \approx 1.5 \times 10^{20}\ \text{Hz} \times 0.6 \times 10^{18}\ \text{s} \approx 10^{38}. \qquad (62)$$

Since we do observe linear polarization in the 200-400 kHz frequency band of GRB 140206A with lower bound of 28 %, this gives a fractional constraint of about $10^{-38}$ on a combination of $\chi$'s. A similar constraint can be obtained for GRB 061122 (the band width times the redshift is about the same). A more detailed modeling may give better limits. The distribution of GRBs is basically isotropic. When this procedure is applied to an ensemble of polarized GRBs from various directions, the relation (20) would be verified to about $10^{-38}$.

Thus, we see that from the pulsar signal propagation, the polarization observations on radio galaxies and the gamma ray burst observations the nonbirefringence condition is verified empirically in spacetime propagarion with accugaracies to $10^{-16}$, $10^{-32}$, and $10^{-38}$. The accuracies of three observational constraints are summarized in Table I. The constitutive tensor can be constructed by the procedure in the proof of the theorem in this subsection to be in the core form (60) with accuracy to $10^{-38}$. Nonbirefringence (no splitting, no retardation) for electromagnetic wave propagation independent of polarization and frequency (energy) is the statement of Galileo Equivalence Principle for photons or WEP I for photons. Hence WEP I for photons in verified to this accuracy in the spacetime propagation.

In the following subsection, we assume (60) [i.e. (28)] is valid and look into the influence of the axion field and the dilaton field of the constitutive tensor on the dispersion relation.

### *3.4. Wave propagation and the dispersion relation in dilaton field and axion field*

We first notice that in the lowest eikonal approximation, the dispersion relation (46) or (47) does not contain the axion piece and does not contain the gradient of fields. Dilaton in (60) goes in this dispersion relation only as an overall scale factor and drops out too.

To derive the influence of the dilaton field and the axion field on the dispersion relation, one needs to keep the second term in equation (31). This has been done for the axion field in references [53, 60, 61, 78-80]. Here we follow the treatment in [63] to develop it for the joint dilaton field and axion field with the constitutive relation (60). Near the origin in a local inertial frame, the constitutive tensor density in dilaton field $\psi$ and axion field $\varphi$ [equation (60)] becomes

$$\chi^{ijkl}(x^m) = [(1/2)\, \eta^{ik}\, \eta^{jl} - (1/2)\, \eta^{il}\, \eta^{kj}]\, \psi(x^m) + \varphi(x^m)\, e^{ijkl} + \text{O}(\delta_{ij}x^i x^j), \qquad (63)$$

where $\eta^{ij}$ is the Minkowski metric with signature $-2$ and $\delta_{ij}$ the Kronecker delta. In the local inertial frame, we use the Minkowski metric and its inverse to raise and lower indices. Substituting (63) into the equation (31) and multiplying by 2, we have

$$\psi A^{i,j}{}_{,j} + \psi A^{j,i}{}_{,j} + \psi_{,j} A^{i,j} - \psi_{,j} A^{j,i} + 2\,\varphi_{,j}\, e^{ijkl}\, A_{k,l} = 0. \tag{64}$$

We notice that (64) is both Lorentz covariant and gauge invariant.

We expand the dilaton field $\psi(x^m)$ and the axion field $\varphi(x^m)$ at the 4-point (event) $P$ with respect to the event (time and position) $P_0$ at the origin as follows:

$$\psi(x^m) = \psi(P_0) + \psi_{,i}(P_0)\, x^i + O(\delta_{ij} x^i x^j), \tag{65a}$$
$$\varphi(x^m) = \varphi(P_0) + \varphi_{,i}(P_0)\, x^i + O(\delta_{ij} x^i x^j). \tag{65b}$$

To look for wave solutions, we use eikonal approximation which does not neglect field gradient/medium inhomogeneity. Choose $z$-axis in the wave propagation direction so that the solution takes the following form:

$$A \equiv (A_0, A_1, A_2, A_3) = (\underline{A}_0, \underline{A}_1, \underline{A}_2, \underline{A}_3)\, e^{ikz - i\omega t} = \underline{A}_i\, e^{ikz - i\omega t}. \tag{66}$$

Expand the solution as

$$A_i = A^{(0)}{}_i + A^{(1)}{}_i + O(2) = [\underline{A}^{(0)}{}_i + \underline{A}^{(1)}{}_i + O(2)]\, e^{ikz - i\omega t} = \underline{A}_i\, e^{ikz - i\omega t}. \tag{67}$$

Now use eikonal approximation to obtain a local dispersion relation. In the eikonal approximation, we only keep terms linear in the derivative of the dilaton field and the axion field; we neglect terms containing the second-order derivatives of the dilaton field or the axion field, terms of $O(\delta_{ij} x^i x^j)$ and terms of mixed second order, e.g. terms of $O(A^{(1)}{}_i x^j)$ or $O(A^{(1)}{}_i \psi_{,j})$; we call all these terms $O(2)$.

Imposing radiation gauge condition in the zeroth order in the weak field/dilute medium approximation, we find to zeroth order, (65) is

$$\psi A^{(0)ij}{}_{,j} = 0,\ \text{or}\ A^{(0)ij}{}_{,j} = 0, \tag{68}$$

and the corresponding zeroth order solution and the dispersion relation are

$$A^{(0)}{}_i = (0, A^{(0)}{}_1, A^{(0)}{}_2, 0) = \underline{A}^{(0)}{}_i\, e^{ikz - i\omega t} = (0, \underline{A}^{(0)}{}_1, \underline{A}^{(0)}{}_2, 0)\, e^{ikz - i\omega t}, \tag{69a}$$

$$\omega = k + O(1). \tag{69b}$$

Substituting (68) and (69a,b) into equation (64), we have

$$\psi A^{(0)ij}{}_{,j} + \psi A^{(1)ij}{}_{,j} + \psi A^{(1)ji}{}_{,j} + \psi_{,j} A^{(0)ij} - \psi_{,j} A^{(0)ji} + 2\,\varphi_{,j}\, e^{ijkl}\, A^{(0)}{}_{k,l} = 0 + O(2). \tag{70}$$

The $i = 0$ and $i = 3$ components of (70) both lead to the same modified Lorentz gauge condition in the dilaton field and the axion field in the O(1) order [63]:

$$A^{(1)j}{}_j = -\psi^{-1}(\psi_{,1} - 2\varphi_{,2})A^{(0)}{}_1 - \psi^{-1}(\psi_{,2} + 2\varphi_{,1})A^{(0)}{}_2 + O(2). \tag{71}$$

Since equation (71) does not contain $\omega$ and $k$, it does not contribute to the determination of the dispersion relation.

Using the gauge condition (71), we obtain the $i = 1$ and $i = 2$ components of equation (70) as

$$(\omega^2 - k^2)\underline{A}^{(0)}{}_1 - ik\underline{A}^{(0)}{}_1\psi^{-1}(\psi_{,0} + \psi_{,3}) - 2ik\underline{A}^{(0)}{}_2\psi^{-1}(\varphi_{,0} + \varphi_{,3}) = 0 + O(2), \tag{72a}$$
$$(\omega^2 - k^2)\underline{A}^{(0)}{}_2 - ik\underline{A}^{(0)}{}_2\psi^{-1}(\psi_{,0} + \psi_{,3}) + 2ik\underline{A}^{(0)}{}_1\psi^{-1}(\varphi_{,0} + \varphi_{,3}) = 0 + O(2). \tag{72b}$$

These two equations determine the dispersion relation in the dilaton field and the axion field:

$$\det \begin{bmatrix} (\omega^2 - k^2) - ik\psi^{-1}(\psi_{,0} + \psi_{,3}) & -2ik\psi^{-1}(\varphi_{,0} + \varphi_{,3}) \\ 2ik\psi^{-1}(\varphi_{,0} + \varphi_{,3}) & (\omega^2 - k^2) - ik\psi^{-1}(\psi_{,0} + \psi_{,3}) \end{bmatrix}$$
$$= [(\omega^2 - k^2) - ik\psi^{-1}(\psi_{,0} + \psi_{,3})]^2 - 4k^2\psi^{-2}(\varphi_{,0} + \varphi_{,3})^2 = 0 + O(2). \tag{73}$$

Its solutions are

$$\omega = k - (i/2)\psi^{-1}(\psi_{,0} + \psi_{,3}) \pm \psi^{-1}(\varphi_{,0} + \varphi_{,3}) + O(2), \text{ or} \tag{74a}$$
$$k = \omega + (i/2)\psi^{-1}(\psi_{,0} + \psi_{,3}) \pm \psi^{-1}(\varphi_{,0} + \varphi_{,3}) + O(2), \tag{74b}$$

with the group velocity $v_g = \partial\omega/\partial k = 1$ independent of polarization. When the dispersion relation is satisfied, (72a) and (72b) have two independent solutions for the polarization eigenvectors $\underline{A}^{(0)}{}_i = (\underline{A}^{(0)}{}_0, \underline{A}^{(0)}{}_1, \underline{A}^{(0)}{}_2, \underline{A}^{(0)}{}_3)$ with

$$\underline{A}^{(0)}{}_1/\underline{A}^{(0)}{}_2 = [2ik\psi^{-1}(\varphi_{,0} + \varphi_{,3})] / [(\omega^2 - k^2) - ik\psi^{-1}(\psi_{,0} + \psi_{,3})]$$
$$= [2ik\psi^{-1}(\varphi_{,0} + \varphi_{,3})] / [\pm 2k\psi^{-1}(\varphi_{,0} + \varphi_{,3})] = \pm i; \tag{75a}$$
$$\underline{A}^{(0)}{}_0 = \underline{A}^{(0)}{}_3 = 0, \tag{75b}$$

for $\omega = k - (i/2)\psi^{-1}(\psi_{,0} + \psi_{,3}) \pm \psi^{-1}(\varphi_{,0} + \varphi_{,3}) + O(2)$ respectively. From (30a), the two polarization eigenstates are left circularly polarized state and right circularly polarized state in varying axion field. This agrees with the electromagnetic wave propagation in axion field as derived earlier [53, 60, 61, 78-80].

With the dispersion (74), the plane-wave solution (66) propagating in the $z$-direction is

$$A \equiv (A_0, A_1, A_2, A_3) = (0, \underline{A}^{(0)}{}_1, \underline{A}^{(0)}{}_2, 0) e^{ikz-i\omega t}$$
$$= (0, \underline{A}^{(0)}{}_1, \underline{A}^{(0)}{}_2, 0) \exp[ikz - ikt \pm (-i)\psi^{-1}(\varphi_{,0} t + \varphi_{,3} z) - (1/2)\psi^{-1}(\psi_{,0} t + \psi_{,3} z)], \tag{76}$$

with $\underline{A}^{(0)}{}_1 = \pm i \underline{A}^{(0)}{}_2$. The additional factor acquired in the propagation is $\exp[\pm (-i) \psi^{-1} (\varphi_{,0} t + \varphi_{,3} z)] \times \exp[-(1/2)\psi^{-1} (\psi_{,0} t + \psi_{,3} z)]$. The first part of this factor, i.e., the axion factor $\exp[\pm (-i) \psi^{-1} (\varphi_{,0} t + \varphi_{,3} z)]$ adds a phase in the propagation. The second part of this factor, i.e., the dilaton factor $\exp[-(1/2) \psi^{-1} (\psi_{,0} t + \psi_{,3} z)]$ amplifies or attenuates the wave according to whether $(\psi_{,0} t + \psi_{,3} z)$ is less than zero or greater than zero. For the right circularly polarized electromagnetic wave, the effect of the axion field in the propagation from a point $P_1 = \{x_{(1)}{}^i\} = \{x_{(1)}{}^0; x_{(1)}{}^\mu\} = \{x_{(1)}{}^0, x_{(1)}{}^1, x_{(1)}{}^2, x_{(1)}{}^3\}$ to another point $P_2 = \{x_{(2)}{}^i\} = \{x_{(2)}{}^0; x_{(2)}{}^\mu\} = \{x_{(2)}{}^0, x_{(2)}{}^1, x_{(2)}{}^2, x_{(2)}{}^3\}$ is to add a phase of $\alpha = \psi^{-1} [\varphi(P_2) - \varphi(P_1)]$ ($\approx \varphi(P_2) - \varphi(P_1)$ for $\psi \approx 1$) to the wave; for left circularly polarized light, the effect is to add an opposite phase [53, 60, 61, 78-80]. Linearly polarized electromagnetic wave is a superposition of circularly polarized waves. Its polarization vector will then rotate by an angle $\alpha$. The effect of the dilaton field is to amplify with a factor $\exp[-(1/2) \psi^{-1} (\psi_{,0} t + \psi_{,3} z)] = \exp[-(1/2) ((\ln \psi)_{,0} t + (\ln \psi)_{,3} z)] = (\psi(P_1)/\psi(P_2))^{1/2}$. The dilaton field contributes to the amplitude of the propagating wave is positive or negative depending on $\psi(P_1)/\psi(P_2) > 1$ or $\psi(P_1)/\psi(P_2) < 1$ respectively.

For plane wave propagating in direction $n^\mu = (n^1, n^2, n^3)$ with $(n^1)^2 + (n^2)^2 + (n^3)^2 = 1$, the solution is

$$A(n^\mu) \equiv (A_0, A_1, A_2, A_3) = (0, \underline{A}_1, \underline{A}_2, \underline{A}_3) \exp(-i kn^\mu x_\mu - i\omega t)$$
$$= (0, \underline{A}_1, \underline{A}_2, \underline{A}_3) \exp[-ikn^\mu x_\mu - ikt \pm(-i)\psi^{-1}(\varphi_{,0}t - n^\mu\varphi_{,\mu}n_\nu x^\nu) - (1/2) \psi^{-1}(\psi_{,0}t + n^\mu\psi_{,\mu}n_\nu x^\nu)], \quad (77)$$

where $\underline{A}_\mu = \underline{A}^{(0)}{}_\mu + n_\mu n^\nu \underline{A}^{(0)}{}_\nu$ with $\underline{A}^{(0)}{}_1 = \pm i \underline{A}^{(0)}{}_2$ and $\underline{A}^{(0)}{}_3 = 0$ [$n_\mu \equiv (-n^1, -n^2, -n^3)$]. There are polarization rotation for linearly polarized light due to axion field gradient, and amplification/attenuation due to dilaton field gradient.

The above analysis is local. In the global situation, choose local inertial frames along the wave trajectory and integrate along the trajectory. Since $\psi$ is a scalar, the integration gives $(\psi(P_1)/\psi(P_2))^{1/2}$ as the amplification factor for the propagation in the dilaton field. For small dilaton field variations, the amplification/attenuation factor is equal to $[1 - (1/2) (\Delta\psi/\psi)]$ to a very good approximation with $\Delta\psi \equiv \psi(P_2) - \psi(P_1)$. Since this factor does not depend on the wave number/frequency and polarization, it will not distort the source spectrum in propagation, but gives an overall amplification/attenuation factor to the spectrum. The axion field contributes to the phase factor and induces polarization rotation as in previous investigations [53, 60, 61, 78-80]. For $\psi \approx 1$ (constant), the induced polarization rotation agrees with previous results which were obtained without considering dilaton effect. If the dilaton field varies significantly, a $\psi$-weight needs to be included in the integration.

The complete agreement with EEP for photon sector requires in addition to Galileo equivalence principle (WEP I; nonbirefringence) for photons: (i) no polarization rotation (WEP II); (iii) no amplification/no attenuation in spacetime propagation; (iii) no spectral distortion. With nonbirefringence, any skewonless spacetime constitutive tensor must be of the form (60), hence no spectral distortion. From (60), (ii) and (iii) implies that the dilaton $\psi$ and axion $\varphi$ must be constant, i.e. no varying dilaton field and no varying axion field; the EEP for photon sector is observed; the spacetime constitutive tensor is of metric-induced form. Thus the three observational conditions are tied to EEP and to metric-induced spacetime constitutive tensor in the photon sector.

In the next subsection, we look into the empirical support of no amplification/no attenuation and no polarization rotation conditions.

### *3.5. No amplification/no attenuation and no polarization rotation constraints on cosmic dilaton field and cosmic axion field*

In this section we look into the observations/experiments to constrain the dilaton field contribution and the axion field contribution to spacetime constitutive tensor density.

*No amplification/no attenuation constraint on the cosmic field*: From equation (76) and (77) in the last section, we have derived that the amplitude and phase factor of propagation in the cosmic dilaton and cosmic axion field is changed by $(\psi(P_1)/\psi(P_2))^{1/2} \times \exp[ikz - ikt \pm (-i)(\varphi(P_1) - \varphi(P_2))t]$. The effect of dilaton field is to give amplification $((\psi(P_1) - \psi(P_2) > 0)$ or attenuation $((\psi(P_1) - \psi(P_2) < 0)$ to the amplitude of the wave independent of frequency and polarization.

The spectrum of the cosmic microwave background (CMB) is well understood to be Planck blackbody spectrum. In the cosmic propagation, this spectrum would be amplified or attenuated by the factor $(\psi(P_1)/\psi(P_2))^{1/2}$. However, the CMB spectrum is measured to agree with the ideal Planck spectrum at temperature $2.7255 \pm 0.0006$ K [81] with a fractional accuracy of $2 \times 10^{-4}$. The spectrum is also red-shifted due to cosmological curvature (or expansion), but this does not change the blackbody character. The measured shape of the CMB spectra does not deviate from Planck spectrum within its experimental accuracy. In the dilaton field the relative increase in power is proportional to the amplitude increase squared, i.e., $\psi(P_1)/\psi(P_2)$. Since the total power of the blackbody radiation is proportional to the temperature to the fourth power $T^4$, the fractional change of the dilaton field since the last scattering surface of the CMB must be less than about $8 \times 10^{-4}$ and we have

$$|\Delta\psi|/\psi \leq 4\,(0.0006/2.7255) \approx 8 \times 10^{-4}. \tag{78}$$

Direct fitting to the CMB data with the addition of the scale factor $\psi(P_1)/\psi(P_2)$ would give a more accurate value.

*Constraints on the cosmic polarization rotation and the cosmic axion field*: From (77), for the right circularly polarized electromagnetic wave, the propagation from a point $P_1$ (4-point) to another point $P_2$ adds a phase of $\alpha = \varphi(P_2) - \varphi(P_1)$ to the wave; for left circularly polarized light, the added phase will be opposite in sign [53]. Linearly polarized electromagnetic wave is a superposition of circularly polarized waves. Its polarization vector will then rotate by an angle $\alpha$. In the global situation, it is the property of (pseudo)scalar field that when we integrate along light (wave) trajectory the total polarization rotation (relative to no $\varphi$-interaction) is again $\alpha = \Delta\varphi = \varphi(P_2) - \varphi(P_1)$ where $\varphi(P_1)$ and $\varphi(P_2)$ are the values of the scalar field at the beginning and end of the wave. The constraints [53, 60, 61, 82-84] listed on the axion field are from the UV polarization observations of radio galaxies and the CMB polarization observations -- 0.02 for Cosmic Polarization Rotation (CPR) mean value $|\langle\alpha\rangle|$ and 0.03 for the CPR fluctuations $\langle(\alpha -$

<$\alpha$>$)^2$>$^{1/2}$.

*Additional constraints to have the unique physical metric*: From (78) the fractional change of dilaton $|\Delta\psi|/\psi$ is less than about $8 \times 10^{-4}$ since the time of the last scattering surface of the CMB. Eötvös-type experiments constrain the fractional variation of dilaton to ~$10^{-10}$ $U$ where $U$ is the dimensionless Newtonian potential in the experimental environment. Vessot-Levine redshift experiment and Hughes-Drever-type experiments give further constraints (section 6 and section 7) [61]. All these constraints are summarized in Table 1. This leads to unique physical metric to high precision for all degrees of freedom except the axion degree of freedom and cosmic dilaton degree of freedom which are only mildly constrained.

### 3.6. *Spacetime constitutive relation including skewons* [62,17]

In this subsection, we review the present status of empirical tests of full local linear spacetime constitutive tensor density (26) of premetric electrodynamics. Since EEP is verified to a good precision, we are mainly concerned with weak EEP violations and weak additional field, i.e. we are assuming $\chi^{(0)ijkl}$ is metric and the components of $\chi^{(1)ijkl}$ are small in most parts of our treatment. We note that *all the formulas in section 3.2 are valid with or without skewonless assumption*.

In particular, *the condition of $^{(Sk)}B_{(1)} = {}^{(P)}B_{(1)} = 0$ and $A_{(1)} = A_{(2)}$ for all directions of wave propagation* still gives (54a-l) without skewonless assumption.

We do not assume skewonless condition in this subsection. The Hehl-Obukhov-Rubilar skewon field (27c) can be represented as

$$^{(Sk)}\chi^{ijkl} = e^{ijmk} S_m{}^l - e^{ijml} S_m{}^k, \tag{79}$$

where $S_m{}^n$ is a traceless tensor with $S_m{}^m = 0$ [6]. From (79), we have

$$^{(Sk)}\chi^{(1)1320} = -S^{(1)}{}_0{}^0 - S^{(1)}{}_2{}^2; \;^{(Sk)}\chi^{(1)1230} = S^{(1)}{}_0{}^0 + S^{(1)}{}_3{}^3; \;^{(Sk)}\chi^{(1)2310} = S^{(1)}{}_0{}^0 + S^{(1)}{}_1{}^1. \tag{80}$$

From $^{(Sk)}$(54g-54l), we must have $^{(Sk)}\chi^{(1)1320} = {}^{(Sk)}\chi^{(1)1230} = {}^{(Sk)}\chi^{(1)2310} = 0$. From (80) and Tr $S_n{}^m = 0$, then all $S^{(1)}{}_0{}^0$, $S^{(1)}{}_1{}^1$, $S^{(1)}{}_2{}^2$ and $S^{(1)}{}_3{}^3$ must vanish.

From (79) together with $^{(Sk)}$(54a-54f), we have

$$S^{(1)}{}_3{}^2 = -S^{(1)}{}_2{}^3; \; S^{(1)}{}_1{}^3 = -S^{(1)}{}_3{}^1; \; S^{(1)}{}_2{}^1 = -S^{(1)}{}_1{}^2; \; S^{(1)}{}_3{}^0 = S^{(1)}{}_0{}^3; \; S^{(1)}{}_1{}^0 = S^{(1)}{}_0{}^1; \; S^{(1)}{}_2{}^0 = S^{(1)}{}_0{}^2. \tag{81}$$

Using the Lorentz metric (*h*-metric in the locally inertia frame) to raise/lower the indices, we have

$$S^{(1)mn} = -S^{(1)nm}, \; S^{(1)}{}_{mn} = -S^{(1)}{}_{nm}. \tag{82}$$

Thus, when $^{(Sk)}$(54a-54l) (9 independent conditions) are satisfied, the skewon degrees of freedom are reduced to 6 (15 − 9) and only Type II skewon field remains.

Under Lorentz (coordinate) transformation, the symmetric part and the anti-symmetric part of $S^{mn}$ transform separately. Hence, with the conditions $^{(Sk)}B = 0$ for all directions of wave propagation, the skewon field is constrained to Type II. The reverse is also true: Since $^{(SkII)}S_{nm}$ is a tensor, when it satisfy $^{(Sk)}B = 0$ for the z-axis of wave propagation, they satisfy $^{(Sk)}B = 0$ for *all directions* of wave propagation. Hence we have the lemma:

*Lemma*: The following three statements are equivalent
(i) $^{(Sk)}B = 0$ for all directions,
(ii) $^{(Sk)}$(54a-54l) hold,
(iii) $^{(Sk)}S_{mn}$ as defined by (79) can be written as $^{(Sk)}S_{mn} = {}^{(SkII)}S_{mn}$ with $^{(SkII)}S_{nm} = - {}^{(SkII)}S_{mn}$.

Proof: (i) → (ii) has been demonstrated in the derivation of $^{(Sk)}$(54a-54l).

(ii) ←→ (iii) has also been demonstrated in the derivation of (80)-(82) and its reversibility.

(iii) → (i). $^{(SkII)}S_{ij}$ is a tensor. If its anti-symmetric property holds in one frame, it holds in any frame. Hence, in any new frame with the propagation in the z-direction, $^{(Sk)}$(54a-54l) hold and we have $^{(Sk)}\underline{B} = 0$ for propagation in the z-direction. Since z-direction can be arbitrary, we have $^{(Sk)}\underline{B} = 0$ for all directions.

The condition of $^{(Sk)}B_{(1)} = {}^{(P)}B_{(1)} = 0$ and $A_{(1)} = A_{(2)}$ for all directions of wave propagation gives (56a,b). Define the anti-symmetric metric $p^{ij}$ as follow:

$$p^{10} \equiv -p^{01} \equiv 2\ ^{(SkII)}\chi^{(1)1220}; p^{20} \equiv -p^{02} \equiv 2\ ^{(SkII)}\chi^{(1)2330}; p^{30} \equiv -p^{03} \equiv 2\ ^{(SkII)}\chi^{(1)3110};$$
$$p^{12} \equiv -p^{21} \equiv 2\ ^{(SkII)}\chi^{(1)1020}; p^{23} \equiv -p^{32} \equiv 2\ ^{(SkII)}\chi^{(1)2030}; p^{31} \equiv -p^{13} \equiv 2\ ^{(SkII)}\chi^{(1)3010};$$
$$p^{00} \equiv p^{11} \equiv p^{22} \equiv p^{33} \equiv 0. \tag{83}$$

It is straightforward to show now that when (54a-l) and (56a-b) are satisfied, then $\chi$ can be written to first-order in terms of the fields $h^{(1)ij}$, $\psi$, $\varphi$, and $p^{ij}$ with $h^{ij} \equiv \eta^{ij} + h^{(1)ij}$ and $h \equiv \det(h_{ij})$ in the following form:

$$\chi^{ijkl} = {}^{(P)}\chi^{(1)ijkl} + {}^{(A)}\chi^{(1)ijkl} + {}^{(SkII)}\chi^{(1)ijkl}$$
$$= \tfrac{1}{2}\ (-h)^{1/2}[h^{ik}h^{jl} - h^{il}h^{kj}]\psi + \varphi e^{ijkl} + \tfrac{1}{2}\ (-\eta)^{1/2}\ (p^{ik}\eta^{jl} - p^{il}\eta^{jk} + \eta^{ik}p^{jl} - \eta^{il}p^{jk}), \tag{84}$$

with

$$^{(P)}\chi^{(1)ijkl} = \tfrac{1}{2}\ (-h)^{1/2}[h^{ik}h^{jl} - h^{il}h^{kj}]\psi, \tag{85a}$$
$$^{(A)}\chi^{(1)ijkl} = \varphi e^{ijkl}, \tag{85b}$$
$$^{(SkII)}\chi^{(1)ijkl} = \tfrac{1}{2}\ (-\eta)^{1/2}\ (p^{ik}\eta^{jl} - p^{il}\eta^{jk} + \eta^{ik}p^{jl} - \eta^{il}p^{jk}). \tag{85c}$$

It is ready to derive the following theorem:

*Theorem*: For linear electrodynamics with skewonful constitutive relation (26) with $^{(Sk)}B = 0$ satisfied for all directions, the following three statements are equivalent to first order

in the field:
(i) $A_{(1)} = A_{(2)}$ and $^{(P)}B = 0$ for all directions, i.e. nonbirefringence in electromagnetic wave propagation,
(ii) (58a-j) hold,
(iii) $\chi^{ijkl}$ can be expressed as (84) with (85a-c).

The proof is similar to that for theorem in subsection 3.3 [62]; readers could readily figure it out.

When the principal part $^{(P)}\chi^{ijkl}$ of the constitutive tensor is induced by metric $h^{ij}$ and dilaton, i.e.

$$^{(P)}\chi^{ijkl} = (-h)^{1/2}[(1/2)h^{ik} h^{jl} - (1/2)h^{il} h^{kj}]\psi, \tag{86}$$

it is easy to check by substitution that

$$A_{(1)} = A_{(2)} \text{ and } {}^{(P)}B_{(1)} = 0. \tag{87}$$

$A_{(1)} = A_{(2)} = {}^{(P)}B = 0$. We have $\xi = -4({}^{(Sk)}B)^2$. There the 3 cases discussed after equation (49) reduce to two cases:

(a) $\xi = 0$, $^{(Sk)}B = 0$: There are no birefringence and no dissipation/amplification in wave propagation;
(b) $\xi < 0$, $^{(Sk)}B \neq 0$: There is no birefringence, but there are both dissipative and amplifying modes in wave propagation.

Now the issue is: When the skewon part of the constitutive tensor is nonzero, what can we say about the spacetime structure empirically?

If $\xi$ is less than zero, i.e. $(A_{(1)} - A_{(2)})^2 + 4({}^{(P)}B)^2 < 4({}^{(Sk)}B)^2$, the dispersion relation (47) is

$$\omega = k [1 + \tfrac{1}{2} (A_{(1)} + A_{(2)}) \pm \tfrac{1}{2} (-\xi)^{1/2} i] + O(2). \tag{88}$$

The exponential factor in the wave solution (36) is of the form

$$\exp(ikz - i\omega t) \sim \exp[ikz - ik (1 + 1/2 (A_{(1)} + A_{(2)})) t] \exp(\pm\tfrac{1}{2} (-\xi)^{1/2} kt). \tag{89}$$

There are both dissipative and amplifying wave propagation modes. In the small $\xi$ limit, the amplification/attenuation factor $\exp(\pm\tfrac{1}{2} (-\xi)^{1/2} kt)$ equals $[1 \pm \tfrac{1}{2} (-\xi)^{1/2} kt]$ to a very good approximation. Since this factor depends on the wave number/frequency, it will distort the source spectrum in propagation.

The spectrum of the cosmic microwave background (CMB) is well understood to be Planck blackbody spectrum. It is measured to agree with the ideal Planck spectrum at temperature $2.7255 \pm 0.0006$ K [81]. The measured shape of the CMB spectra does not deviate from Planck spectrum within its experimental accuracy. The agreement for the overall shape with a fit to Planck plus a linear factor $[1 \pm \tfrac{1}{2}(-\xi)^{1/2}kt]$ is to agree with Planck to better than $10^{-4}$. Planck Surveyor has nine bands of detection from 30 to 857 GHz [86]. For weak propagation deviation, the amplitude of the wave is increased or decreased linearly as $\tfrac{1}{2}(-\xi)^{1/2}kt$ depending on frequency. For cosmic propagation, the CMB amplitude change due to redshift (or blue shift) is universal. The frequency (wave number) change is proportional to $(1 + z(t))$ with $z(t)$ the redshift factor at time $t$ of propagation. We need to replace $kt$ in the $[1\pm \tfrac{1}{2}(-\xi)^{1/2}kt]$ factor by the integral

$$\int k(t)\, dt = \int k(t_0)(1+z(t))\, dt \equiv (1+ <z(t)>)\, k(t_0)\,(t_0 - t_1), \qquad (90)$$

with $<z(t)>$ the average of $z(t)$ during propagation defined by the last equality of (90), $t_0$ the present time (the age of our universe), and $t_1$ the time at the photon decoupling epoch. According to *Planck* 2013 results [85], the age of our universe $t_0$ is 13.8 Gyr, the decoupling time $t_1$ is 0.00038 Gyr, hence $(t_0 - t_1)$ is ~13.8 Gyr, and $z(t_1)$ is 1090. Using *Planck* $\Lambda$CDM concordance model, the factor $(1+ <z(t)>)$ is estimated to be about 3 and the value $(1+ <z(t)>)(t_0 - t_1)$ is more than 40 Gyr. The factor $(1+ <z(t)>)$ multiply by $(t_0 - t_1)$ is the angular diameter distance $D_A$ at which we are observing the CMB and is equal to the comoving size of the sound horizon at the time of last-scattering, $r_s(z(t_1))$, divided by the observed angular size $\theta_* = r_s/D_A$ from seven acoustic peaks in the CMB anisotropy spectrum. From Planck results, $r_s = 144.75 \pm 0.66$ Mpc and $\theta_* = (1.04148 \pm 0.00066) \times 10^{-2}$. Hence, we have $D_A = r_s/\theta_* = 13898 \pm 64$ Mpc $= 45.328 \pm 0.21$ Gyr. This is consistent with our integral estimation.

For the highest frequency band $\omega$ is $2\pi \times 857$ GHz. The amplification/dissipation in fraction is

$$\tfrac{1}{2}(-\xi)^{1/2} k \times 45.328\text{ Gyr} = 3.8 \times 10^{30}(-\xi)^{1/2}. \qquad (91)$$

For the lowest frequency band $\omega$ is $2\pi \times 30$ GHz; the effect is about $\pm 3.5$ % of (91). From CMB observations that the spectrum is less than $10^{-4}$ deviation, we have

$$(-\xi)^{1/2} < 2.6 \times 10^{-35}. \qquad (92)$$

When the spacetime constitutive tensor is constructed from metric, dilaton and axion plus skewon, the principal part $^{(P)}\chi^{ijkl}$ of the constitutive tensor is given by (86), there are two cases, (a) $^{(Sk)}B = 0$ and (b) $^{(Sk)}B \neq 0$ as mentioned after Eq. (87). For case (a) $\xi = 0$, there are no birefringence and no dissipation/amplification in wave propagation; by the *Theorem* in this subsection, the skewon part must be of Type II. For case (b) $\xi < 0$, $^{(Sk)}B \neq 0$, there are both dissipative and amplifying modes in wave propagation and we can apply the (92) from the CMB observations to constrain the skewon part of the constitutive tensor as follows

$$\tfrac{1}{2}(-\xi)^{1/2} = |^{(Sk)}B| = \tfrac{1}{2}|(B_{(1)} - B_{(2)})| = |^{(Sk)}\chi^{(1)1020} + {}^{(Sk)}\chi^{(1)1323} - {}^{(Sk)}\chi^{(1)1023} - {}^{(Sk)}\chi^{(1)1320}| < 1.3\times 10^{-35}, \quad (93)$$

for propagation in the $z$-direction. Since the CMB observation is omnidirectional, we have the above constraint for many directions. From a few superpositions, we obtain the *Lemma* in this subsection, hence the constraints (54a-l) hold to ~ a few × $20^{-35}$ and the spacetime skewon field is Type II with type I skewon field constrained to ~ a few × $20^{-35}$ cosmologically in the first order. Thus, *the significant skewon field must be of Type II with six degrees of freedom in the first order.*

*Constraints on the skewon field in the second order* [17]

For metric principal part plus skewon part, we have shown that the Type I skewon part is constrained to < a few × $10^{-35}$ in the weak field/weak EEP violation limit. Type II skewon part is not constrained in the first order. In the second order Obukhov and Hehl have shown in Sec.IV.A.1 of [86] that it induces birefringence; since the nonbirefringence observations are precise to $10^{-38}$ as listed in Table I, they constrain the Type II skewon part to ~ $10^{-19}$ [17, 86]. However, an additional nonmetric induced second-order contribution to the principal part constitutive tensor compensates the Type II skewon birefringence and makes it nonbirefringent [17]. This second-order contribution is just the extra piece to the (symmetric) core-metric principal constitutive tensor induced by the antisymmetric part of the asymmetric metric tensor $q^{ij}$ [17]. Table 3 lists various 1st-order and 2nd-order effects in wave propagation on media with the core-metric based constitutive tensors [17]. In the following subsection, we review the spacetime/medium with constitutive tensor induced from asymmetric metric.

Table 3. Various 1st-order and 2nd-order effects in wave propagation on media with the core-metric based constitutive tensors. $^{(P)}\chi^{(c)}$ is the extra contribution due to antisymmetric part of asymmetric metric to the core-metric principal part for canceling the skewon contribution to birefringence/amplification-dissipation [17].

| Constitutive tensor | Birefringence (in the geometric optics approximation) | Dissipation/ amplification | Spectro-scopic distortion | Cosmic polarization rotation |
|---|---|---|---|---|
| Metric: ½ $(-h)^{1/2}[h^{ik} h^{jl} - h^{il} h^{kj}]$ | No | No | No | No |
| Metric + dilaton: ½ $(-h)^{1/2}[h^{ik} h^{jl} - h^{il} h^{kj}]\psi$ | No (to all orders in the field) | Yes (due to dilaton gradient) | No | No |
| Metric + axion: ½ $(-h)^{1/2}[h^{ik} h^{jl} - h^{il} h^{kj}] + \varphi e^{ijkl}$ | No (to all orders in the field) | No | No | Yes (due to axion gradient) |
| Metric + dilaton + axion: ½ $(-h)^{1/2}[h^{ik} h^{jl} - h^{il} h^{kj}]\psi + \varphi e^{ijkl}$ | No (to all orders in the field) | Yes (due to dilaton gradient) | No | Yes (due to axion gradient) |
| Metric + type I skewon | No to first order | Yes | Yes | No |
| Metric + type II skewon | No to first order; yes to 2nd order | No to first order and to 2nd order | No | No |
| Metric + $^{(P)}\chi^{(c)}$ + type II skewon | No to first order; no to 2nd order | No to first order and to 2nd order | No | No |
| Asymmetric metric induced: ½ $(-q)^{1/2}(q^{ik}q^{jl} - q^{il}q^{jk})$ | No (to all orders in the field) | No | No | Yes (due to axion gradient) |

### 3.7. Constitutive tensor from asymmetric metric and Fresnel equation

Eddington [87], Einstein & Straus [88], and Schrödinger [89, 90] considered asymmetric metric in their exploration of gravity theories. Just like we can build spacetime constitutive tensor from the (symmetric) metric as in metric theories of gravity, we can also build it from the asymmetric metric. Let $q^{ij}$ be the asymmetric metric as follows:

$$\chi^{ijkl} = \tfrac{1}{2} (-q)^{1/2}(q^{ik}q^{jl} - q^{il}q^{jk}), \tag{94}$$

with $q = \det^{-1}(^{(S)}q^{ij})$. When $q^{ij}$ is symmetric, this definition reduces to that of the metric theories of gravity. The constitutive law (94) was also put forward by Lindell and Wallen [91] as Q-medium. Resolving the asymmetric metric into symmetric part $^{(S)}q^{ij}$ and antisymmetric part $^{(A)}q^{ij}$:

$$q^{ij} = {}^{(S)}q^{ij} + {}^{(A)}q^{ij}, \text{ with } {}^{(S)}q^{ij} \equiv \tfrac{1}{2} (q^{ij} + q^{ji}) \text{ and } {}^{(A)}q^{ij} \equiv \tfrac{1}{2} (q^{ij} - q^{ji}), \tag{95}$$

we can decompose the constitutive tensor into the principal part $^{(P)}\chi^{ijkl}$, the axion part $^{(Ax)}\chi^{ijkl}$ and skewon part $^{(Sk)}\chi^{ijkl}$ as follows [62,92]:

$$\chi^{ijkl} = \tfrac{1}{2} (-q)^{1/2}(q^{ik}q^{jl} - q^{il}q^{jk}) = {}^{(P)}\chi^{ijkl} + {}^{(Ax)}\chi^{ijkl} + {}^{(Sk)}\chi^{ijkl}, \tag{96a}$$

with

$$^{(P)}\chi^{ijkl} \equiv \tfrac{1}{2} (-q)^{1/2} ({}^{(S)}q^{ik}\,{}^{(S)}q^{jl} - {}^{(S)}q^{il}\,{}^{(S)}q^{jk} + {}^{(A)}q^{ik}\,{}^{(A)}q^{jl} - {}^{(A)}q^{il}\,{}^{(A)}q^{jk} - 2\,{}^{(A)}q^{[ik}\,{}^{(A)}q^{jl]}), \tag{96b}$$

$$^{(Ax)}\chi^{ijkl} \equiv (-q)^{1/2}\,{}^{(A)}q^{[ik}\,{}^{(A)}q^{jl]}, \tag{96c}$$

$$^{(Sk)}\chi^{ijkl} \equiv \tfrac{1}{2} (-q)^{1/2} ({}^{(A)}q^{ik}\,{}^{(S)}q^{jl} - {}^{(A)}q^{il}\,{}^{(S)}q^{jk} + {}^{(S)}q^{ik}\,{}^{(A)}q^{jl} - {}^{(S)}q^{il}\,{}^{(A)}q^{jk}). \tag{96d}$$

The axion part $^{(Ax)}\chi^{ijkl}$ only comes from the second order terms of $^{(A)}q^{il}$.

Using $^{(S)}q^{ij}$ to raise and its inverse to lower the indices, we have as equation (16) in [62]

$$S_i{}^j = \tfrac{1}{4}\, \varepsilon_{ijmk}\,{}^{(A)}q^{mk};\;{}^{(A)}q^{mk} = -\,\varepsilon^{mkij} S_{ij}, \tag{97}$$

where $\varepsilon_{ijmk}$ and $\varepsilon^{mkij}$ are respectively the completely antisymmetric covariant and contravariant tensors with $\varepsilon^{0123} = 1$ and $\varepsilon_{0123} = -1$ in local inertial frame. Thus the skewon field $S_{ij}$ from asymmetric metric $q^{ik}$ is antisymmetric and is of Type II.

*Dispersion relation in the geometrical optics limit.* The dispersion relation for the wave covector $q_i$ of electromagnetic propagation with general constitutive tensor (26) in the geometric-optics limit is given by *the generalized covariant Fresnel equation* [6]:

$$G^{ijkl}(\chi)q_i q_j q_k q_l = 0, \tag{98}$$

where $G^{ijkl}(\chi)\,(= G^{(ijkl)}(\chi))$ is a completely symmetric fourth order Tamm-Rubilar (TR) tensor density of weight $+1$ defined by

$$G^{ijkl}(\chi) \equiv (1/4!)\,\underline{e}_{mnpq}\,\underline{e}_{rstu}\,\chi^{mnr(i}\chi^{j|ps|k}\chi^{l)qtu}. \tag{99}$$

There are two ways to obtain the Tamm-Rubilar tensor density (99) for the dispersion relation (98). One way is by straightforward calculation; the other is by covariant method [92]. In the Appendix of Ref. [43], we outline the straightforward calculation to obtain the Tamm-Rubilar tensor density $G^{ijkl}(\chi)$ for the asymmetric metric induced constitutive tensor:

$$G^{ijkl}(\chi) = (1/8)\,(-q)^{3/2}\det(q^{ij})\,q^{(ij}q^{kl)} = (1/8)\,(-q)^{3/2}\det(q^{ij})\,{}^{(S)}q^{(ij\,(S)}q^{kl)}. \tag{100}$$

Except for a scalar factor, (100) is the same as for metric-induced constitutive tensor with ${}^{(S)}q_{ij}$ replacing the metric $g_{ij}$ or $h_{ij}$. Therefore in the geometric optical approximation, there is no birefringence and the unique light cone is given by the metric ${}^{(S)}q_{ij}$.

*Constraints on asymmetric-metric induced constitutive tensor* [17]. Although the asymmetric-metric induced constitutive tensor leads to a Fresnel equation which is nonbirefringent, it contains an axionic part:

$${}^{(Ax)}\chi^{ijkl} \equiv (-q)^{1/2\,(A)}q^{[ik\,(A)}q^{jl]} = \varphi\,e^{ijkl};\ \varphi \equiv (1/4!)\,e_{ijkl}\,(-q)^{1/2\,(A)}q^{[ik\,(A)}q^{jl]}, \tag{101}$$

which induces polarization rotation in wave propagation. Constraints on CPR and its fluctuation limit the axionic part and therefore also constrain the asymmetric metric. The variation of $\varphi$ ($\equiv (1/4!)\,e_{ijkl}\,(-q)^{1/2\,(A)}q^{[ik\,(A)}q^{jl]}$) is limited by observations [82-84,60,61] on the cosmic polarization rotation to $< 0.02$ and its fluctuation to $< 0.03$ since the last scattering surface, and in turn constrains the antisymmetric metric of the spacetime for this degree of freedom. The antisymmetric metric has 6 degrees of freedom. Further study of the remaining 5 degrees of freedom experimentally to find either evidence or more constraints would be desired.

Theoretically, there are two issues: one is whether the asymmetric-metric induced constitutive tensors with additional axion piece are the most general nonbirefringent media in the lowest geometric optics limit; the other is what they play in the spacetime structure and in the cosmos.

### 3.8. *Empirical foundation of the closure relation for skewonless case* [17,62]

There are two equivalent definitions of constitutive tensor which are useful in various discussions (see, e. g., Hehl and Obukhov [6]). The first one is to take a dual on the first 2 indices of $\chi^{ijkl}$:

$$\kappa_{ij}{}^{kl} \equiv (1/2)\underline{e}_{ijmn}\,\chi^{mnkl}, \tag{102}$$

where $\underline{e}_{ijmn}$ is the completely antisymmetric tensor density of weight $-1$ with $\underline{e}_{0123} = 1$. Since $e_{ijmn}$ is a tensor density of weight $-1$ and $\chi^{mnkl}$ a tensor density of weight $+1$, $\kappa_{ij}{}^{kl}$ is a (twisted) tensor. From (102), we have

$$\chi^{mnkl} = (1/2)e^{ijmn}\kappa_{ij}{}^{kl}. \tag{103}$$

With this definition of constitutive tensor $\kappa_{ij}{}^{kl}$, the constitutive relation (12) becomes

$$*H_{ij} = \kappa_{ij}{}^{kl} F_{kl}, \tag{104}$$

where $*H_{ij}$ is the dual of $H^{ij}$, i.e.

$$*H_{ij} \equiv (1/2)\, \underline{e}_{ijmn} H^{mn}. \tag{105}$$

The second equivalent definition of the constitutive tensor is to use a $6 \times 6$ matrix representation $\kappa_I{}^J$. Since $\kappa_{ij}{}^{kl}$ is nonzero only when the antisymmetric pairs of indices $(ij)$ and $(kl)$ have values (01), (02), (03), (23), (31), (12), these index pairs can be enumerated by capital letters $I, J, \ldots$ from 1 to 6 to obtain $\kappa_I{}^J (\equiv \kappa_{ij}{}^{kl})$. With the relabeling, $F_{ij} \rightarrow F_I$, $H^{ij} \rightarrow H^I$, $\underline{e}_{ijmn} \rightarrow \underline{e}_{IJ}$, $e^{ijmn} \rightarrow e^{IJ}$. We have $F_I = (\boldsymbol{E}, -\boldsymbol{B})$ and $(*H)_I = (-\boldsymbol{H}, -\boldsymbol{D})$. $\underline{e}_{IJ}$ and $e^{IJ}$ can be expressed in matrix form as

$$\underline{e}_{IJ} = e^{IJ} = \begin{pmatrix} 0 & \boldsymbol{I}_3 \\ \boldsymbol{I}_3 & 0 \end{pmatrix}, \tag{106}$$

where $\boldsymbol{I}_3$ is the $3 \times 3$ unit matrix. In terms of this definition, the constitutive relation (104) becomes

$$*H_I = 2\, \kappa_I{}^J F_J, \tag{107}$$

where $*H_I \equiv *H_{ij} = e_{IJ} H^J$. The axion part $^{(\text{Ax})}\chi^{ijkl} (= \varphi\, e^{ijkl})$ now corresponds to

$$^{(\text{Ax})}\kappa_I{}^J = \varphi \begin{pmatrix} \boldsymbol{I}_3 & 0 \\ 0 & \boldsymbol{I}_3 \end{pmatrix} = \varphi\, \boldsymbol{I}_6, \tag{108}$$

where $\boldsymbol{I}_6$ is the $6 \times 6$ unit matrix. The principal part and the axion part of the constitutive tensor all satisfy the following equation (the skewonless condition):

$$e^{KJ}\kappa_J{}^I = e^{IJ}\kappa_J{}^K. \tag{109}$$

In terms of $\kappa_{ij}{}^{kl}$ and re-indexed $\kappa_I{}^J$, the constitutive tensor (60) is represented in the following forms:

$$\kappa_{ij}{}^{kl} = (1/2)\, \underline{e}_{ijmn}\, \chi^{mnkl} = (1/2)\, \underline{e}_{ijmn}\, (-h)^{1/2}\, h^{mk}\, h^{nl}\, \psi + \varphi\, \delta_{ij}{}^{kl}, \tag{110}$$

$$\kappa_I{}^J = (1/2)\, \underline{e}_{ijmn}\, (-h)^{1/2}\, h^{mk}\, h^{nl}\, \psi + \varphi\, \delta_I{}^J, \tag{111}$$

where $\delta_{ij}{}^{kl}$ is a generalized Kronecker delta defined as

$$\delta_{ij}{}^{kl} = \delta_i^k \, \delta_j^l - \delta_i^l \, \delta_j^k. \tag{112}$$

In the derivation, we have used the formula

$$\underline{e}_{ijmn} \, e^{mnkl} = 2 \, \delta_{ij}{}^{kl}. \tag{113}$$

Let us calculate $\kappa_{ij}{}^{kl}\kappa_{kl}{}^{pq}$ for the constitutive tensor (110):

$$\begin{aligned}
\kappa_{ij}{}^{kl} \, \kappa_{kl}{}^{pq} &= [(1/2) \, \underline{e}_{ijmn} \, (-h)^{1/2} \, h^{mk} \, h^{nl} \, \psi + \varphi \, \delta_{ij}{}^{kl}] \, [(1/2) \, \underline{e}_{klrs} \, (-h)^{1/2} \, h^{rp} \, h^{sq} \, \psi + \varphi \, \delta_{kl}{}^{pq}] \\
&= -(1/2) \, \delta_{ij}{}^{pq}\psi^2 + 2 \, \delta_{ij}{}^{pq}\varphi^2 + 2 \, \underline{e}_{ijrs} \, (-h)^{1/2} \, h^{rp} \, h^{sq} \, \varphi \, \psi \\
&= -(1/2) \, \delta_{ij}{}^{pq}\psi^2 + 4 \, \varphi \, {}^{(P)}\kappa_{ij}{}^{pq} - 2 \, \delta_{ij}{}^{pq} \, \varphi^2,
\end{aligned} \tag{114}$$

where we have used (113) and the following relations

$$e_{klrs} \, h^{mk} \, h^{nl} \, h^{rp} \, h^{sq} = e^{mnpq} \det(h^{uv}), \tag{115}$$
$$\det(h^{uv}) = [\det(h_{uv})]^{-1} = h^{-1}, \tag{116}$$
$$\delta_{ij}{}^{kl} \, \delta_{kl}{}^{pq} = 2 \, \delta_{ij}{}^{pq}. \tag{117}$$

In terms of the six-dimensional index $I$, equation (114) becomes

$$\kappa_I{}^J \, \kappa_J{}^K = (1/2)\kappa_{ij}{}^{kl} \, \kappa_{kl}{}^{pq} = -(1/4)\psi^2 \delta_I{}^K + 2^{(P)}\kappa_I{}^K \, \varphi - \delta_{ij}{}^{pq} \, \varphi^2 = -(1/4)\psi^2 \, \delta_I{}^K + 2^{(P)}\kappa_I{}^K \, \varphi - \delta_I{}^K \, \varphi^2. \tag{118}$$

Thus the matrix multiplication of $\kappa_I{}^J$ with itself is a linear combination of itself and the identity matrix, and generates a closed algebra of linear dimension 2. The algebraic relation (118) is a closure relation that generalizes the following closure relation in electrodynamics:

$$\kappa \, \kappa = (\kappa_I{}^J \, \kappa_J{}^K) = (1/6) \, \mathrm{tr}(\kappa \, \kappa) \, \mathbf{I}_6. \tag{119}$$

The matrix multiplication of $\kappa_I{}^J$ satisfies the closure relation (119). In case $\varphi = 0$, the axion part $^{(Ax)}\kappa_I{}^J$ of the constitutive tensor vanishes and (118) reduces to the closure relation (119).

    From the nonbirefringence condition (60), we derive the closure relation (118) in a number of algebraic steps which consist of order 100 individual operations of addition/subtraction or multiplication. Equation (60) is empirically verified to $10^{-38}$. Therefore equation (118) is empirically verified to $10^{-37}$ (precision $10^{-38}$ times $100^{1/2}$). Hence, when there are no axion and no dilaton, the closure relation (119) is empirically verified to $10^{-37}$. For dilaton is constrained to $8 \times 10^{-4}$, if one allow for dilaton, relation (119) is verified to $8 \times 10^{-4}$ since the last scattering surface of CMB; for axion is constrained to $10^{-2}$, if one allow for axion in addition, relation (119) is verified to $10^{-2}$ since the last scattering surface of CMB. As pointed out by Favaro (private communication), the above method could also readily applied to the other 3 variants of closure relations (equations (3.2), (3.3), (3.4) in [92]).

    The closure relation (119) can also be called idempotent condition for it states that

the multiplication of $\kappa$ by itself goes back essentially to itself. Toupin [93], Schonberg [94], and Jadczyk [95] in their theoretical approach started from this condition to obtain metric induced constitutive tensor with a dilaton degree of freedom. In this section, we have started with Galileo equivalence principle for photons, i.e. the nonbirefringence condition, to obtain the metric induced core metric form with a dilaton degree of freedom and an axion degree of freedom for the constitutive tensor and then the generalized closure relation (118). We have also shown that (118) is verified empirically to very high precision. Thus in the axionless (and skewonless) case, the birefringence condition and idempotent condition are equivalent and both are verified to empirically to high precision.

## 4. From Galileo equivalence principle to Einstein equivalence principle (EEP)

In section 3, we have used equivalence principles in the photon sector to constrain the gravitational coupling to electromagnetism and the structure of spacetime from premetric electrodynamics. In this section, we review and discuss theoretically to what extent Galileo equivalence principle leads to Einstein equivalence principle, i.e. Schiff's conjecture.

In 1970s, we used Galileo Equivalence Principle and derived its consequences for an electromagnetic system with Lagrangian density $L$ $(= L_\text{I}^{(\text{EM})} + L_\text{I}^{(\text{EM-P})} + L_\text{I}^{(\text{P})})$ where the electromagnetic field Lagrangian $L_\text{I}^{(\text{EM})}$ and the field-current interaction Lagrangian $L_\text{I}^{(\text{EM-P})}$ are given by (17a,b), and the particle Lagangian $L_\text{I}^{(\text{P})}$ is given by $-\Sigma_I m_I (ds_I)/(dt) \delta(\boldsymbol{x}-\boldsymbol{x}_I)$ with $m_I$ the mass of the $I$th particle, $s_I$ its 4-line element from the metric $g_{ij}$, $\boldsymbol{x}_I$ its position 3-vector, $\boldsymbol{x}$ the coordinate 3-vector, and $t$ the time coordinate [20, 21]:

$$L = L_\text{I}^{(\text{EM})} + L_\text{I}^{(\text{EM-P})} + L_\text{I}^{(\text{P})} = -(1/(16\pi))\chi^{ijkl} F_{ij} F_{kl} - A_k J^k - \Sigma_I m_I (ds_I)/(dt) \delta(\boldsymbol{x}-\boldsymbol{x}_I), \quad (120)$$

$$J^k = \Sigma_I e_I (d\boldsymbol{x}_I)^k/(dt) \delta(\boldsymbol{x}-\boldsymbol{x}_I). \quad (120a)$$

Here $e_I$ is the charge of the $I$th particle. In (120), only the part of $\chi^{ijkl}$ which is symmetric under the interchange of index pairs $ij$ and $kl$ contributes to the Lagrangian, i.e. the constitutive tensor is effectively skewonless. This framework is termed $\chi$-$g$ framework.

The result of imposing Galileo Equivalence Principle is that the constitutive tensor density $\chi^{ijkl}$ can be constrained and expressed in metric form with additional pseudoscalar (axion) field $\varphi$:

$$\chi^{ijkl} = (-g)^{1/2}[(1/2)g^{ik} g^{jl} - (1/2)g^{il} g^{kj}] + \varphi e^{ijkl}, \quad (121)$$

where $g^{ij}$ is the metric of the geodesic motions of particles, $g_{ij}$ is the inverse of $g^{ij}$, $g = \det(g_{ij})$, and $e^{ijkl}$ is the completely anti-symmetric tensor density with $e^{0123} = 1$ as defined in section 3. Hence the metric $g^{ij}$ generates the light cone for electromagnetic wave propagation also. The constraint (121) dictates the gravity coupling to electromagnetic field to be metric plus one additional axionic freedom. With this one axionic freedom the EEP is violated, and therefore the Schiff's conjecture is invalid. However, the spirit of Schiff's conjecture is useful and constrains the gravity coupling effectively. Since the theory with constitutive tensor does not obey EEP, it is a nonmetric theory.

The theory with $\varphi \neq 0$ is a pseudoscalar theory with important astrophysical and cosmological consequences. Its effect on electromagnetic wave propagation is that the polarization rotation of linearly polarized light is proportional to the difference of the (pseudo)scalar field at the two end points. We have discussed this in detail in section 3.4 and use CPR observations to constrain it. This is an example that investigations in fundamental physical laws lead to implications in cosmology. Investigations of CP problems in high energy physics leads to a theory with a similar piece of Lagrangian with φ the axion field for QCD [96-103].

In the nonmetric theory with $\chi^{ijkl}$ ($\varphi \neq 0$) given by Eq. (121) [20, 21, 40, 53], there are anomalous torques on electromagnetic-energy-polarized bodies so that different test bodies will change their rotation state differently, like magnets in magnetic fields. Since the motion of a macroscopic test body is determined not only by its trajectory but also by its rotation state, the motion of polarized test bodies will not be the same. We, therefore, have proposed the following stronger weak equivalence principle (WEP II) to be tested by experiments, which states that in a gravitational field, both the translational and rotational motion of a test body with a given initial motion state is independent of its internal structure and composition (universality of free-fall motion) (Section 2.2) [20, 21]. To put in another way, the behavior of motion including rotation is that in a local inertial frame for test-bodies. If WEP II is violated, then EEP is violated. Therefore from above, in the $\chi$-$g$ framework, the imposition of WEP II guarantees that EEP is valid. These are the reasons for us to propose WEP II. The $\chi$-$g$ framework has been extended to nonabelian gauge fields for studying the interrelations of equivalence principles with similar conclusions [104].

From the empirical side, WEP I for unpolarized bodies is verified to very high precision. However, these experiments only constrain 2 degrees of freedom of $\chi$'s for connecting with gravity coupling of matter. To constrain and connect more degrees of freedom of $\chi$'s to gravity coupling of matter, we propose to perform WEP experiments on various polarized test-bodies in 1970s – both electromagnetic polarized and spin polarized test bodies. These polarized experiments are also crucial to probe the role of spin and polarization in gravity. Now with the spacetime constitutive tensor density constrained to the core metric form (60) to ultra-precision $10^{-38}$, the polarized WEP experiments will test the gravity-matter interaction more than gravity-radiation interaction. In Sec. 7, we will update our review [61] on the search for the long range/intermediate range spin-spin, spin-monopole and spin-cosmic interactions.

## 5. EEP and Universal Metrology

EEP states that all the local physics is the same everywhere at any time in our cosmos. Therefore if we base our metrology everywhere at anytime on local physics with a universal procedure, we have a universal metrology (see, e,g. Ni [105], Petley [106]). For metrology, we need unit standards. At present all basic standards except for the prototype mass standard are based on physical laws, their fundamental constants and the microscopic properties of matter. The Einstein Equivalence Principle (EEP) says, in essence, local physics is the same everywhere. Therefore, to the precision of its empirical tests, EEP warrants the universality of these standards and their implementations.

The name Système International d'Unités (International System of Units), with the abbreviation SI, was adopted by the 11$^{th}$ Conférence Générale des Poids et Mesures in 1960. After 1983 redefinition of meter as the length of path traveled by light in a vacuum during a time interval of 1/299792458 of a second, all definition of SI units can be traced to the definition of second and kilogram. The second is defined as the duration of 9 192 631 770 periods of the radiation corresponding to the transition between the two hyperfine levels of the ground state of the cesium-133 atom. The kilogram is the unit of mass; it is equal to the mass of the international prototype of the kilogram [a cylinder of platinum-iridium] (IPK). IPK is the only physical artefact in the definition of SI 7 base units (second, meter, kilogram, ampere, kelvin, mole and candela for 7 base quantities time, length, mass, electric current, thermodynamic temperature, amount of substance and luminous intensity respectively). Although the uncertainty of the mass of IPK is zero by convention, there are evidence that the mass of IPK varies with a fraction of the order of $10^{-8}$ after storage or cleaning with the estimated relative instability $\delta m/m \approx 5 \times 10^{-8}$ over the past 100 years [107]. When the mass unit is redefined by natural invariants, the SI system will be free of artefacts. In order to ensure continuity of mass metrology, it has been agreed that the relative uncertainty of any new realization must be less than $2 \times 10^{-8}$ (See, e.g. Becker [108]). Sanchez et al. [109] in National Research Council of Canada determined the Planck's constant $h$ using the watt balance to be $6.62607034(12) \times 10^{-34}$ J s within $2 \times 10^{-8}$ relative uncertainty. NIST has reached $5 \times 10^{-8}$ relative uncertainty and is building a new watt balance to reach $2 \times 10^{-8}$ relative uncertainty [110]. The silicon sphere experiment of counting atoms to determine the Avogadro constant reached $3 \times 10^{-8}$ relative uncertainty (See, e.g., Becker [108]). In 2014, the Avogadro constant $N_A$ and derived Planck constant $h$ based on the absolute silicon molar mass measurements with their standard uncertainties are $6.02214076(19) \times 10^{23}$ mol$^{-1}$ and $6.62607017(21) \times 10^{-34}$ J s [111]. The three measurements of NIST [111], PTB [112], and NMIJ [113] agree within their stated uncertainties and also agree with the NRC watt balance measurement with 1 σ. These experimental progresses set the stage for a new definition of kilogram using Planck constant/Avogadro number. Time is becoming mature to replace all the definitions of units using natural invariants.

In 2018, the 5 SI base quantities -- time, length, mass, electric current, and thermodynamic temperature -- will be replaced by frequency, velocity, action, electric charge, and heat capacity, pending upon the expected final resolution of the 26$^{th}$ Conférence Générale des Poinds et Mesures (CGPM) (See, e.g., [110]). The two defining constants for frequency and velocity will be the same as the present SI defining constants of time and length. The defining constants for action, electric charge, heat capacity, and amount of substance will be the Planck constant $h$, the elementary charge $e$, the Boltzmann constant $k$ and the Avogadro constant $N_A$ respectively. The mass unit can be traced to action unit defined by the Planck constant using watt balance or to amount of substance defined by the Avagadro constant based on counting the atoms in a $^{28}$Si crystal. In 2018, both methods should reach an uncertainty smaller than $2 \times 10^{-8}$ to guarantee consistency and continuity. The relative uncertainty of $N_A h$ at present is $7 \times 10^{-10}$ (CODATA 2010 adjustment [114]) to guarantee consistency at the $2 \times 10^{-8}$ level.

With the new definition of units based on physical invariants of nature, the applicability becomes wider; as long as the physical laws which the units are based are valid, the standards and metrology are universal. In section 3, we have seen that the unique light cone is experimental verified to $10^{-38}$ via γ-ray observations at cosmological distance; it verifies the Galileo equivalence principle for photons/electromagnetic wave packets to this accuracy. This constrains the spacetime (vacuum) constitutive tensor to core metric form with additional dilaton and axion degrees of freedom. In the solar system the varational of the dilaton field is constrained to $10^{-10}$ $U$; in the cosmos, the dilaton field is constrained to $8 \times 10^{-4}$ (Table 1). The universal metrology system is truly universal with the present accuracies. In case the accuracies are pushed further, we either verify equivalences principles further or discover new physics. Thus we see that universal metrology and equivalence principles go hand-in hand.

Equivalence principles play very important roles both in the Newtonian theory of gravity and relativistic theories of gravity. The ranges of validity of these equivalence principles or their possible violations give clues and/or constraints to the microscopic origins of gravity. They will be even more important when the precisions of the tests become higher. To pursue further tests of EEP, we have to look into precise experiments and observations in our laboratory, in the solar system, and in diverse astrophysical and cosmological situations. All of these depend on the progress in the field of precision measurement, and demands more precise standards. The constancy of constants is implied by equivalence principles. Their variations give new physics.

The frequency measurement has the best relative uncertainty at present. The optical clocks are reaching relative uncertainties at the $10^{-18}$ level [115]. When the comparison of optical clocks becomes common, it is anticipated that the frequency stardards will go optical. Further improvement in the frequency measurements will have profound impact on precision measurement and gravity experiment. In the realm of gravitational wave detection, the influence will be to enhance the Doppler tracking method and the PTA method [116]. An array of clocks may even become an alternate method for detecting low frequency gravitational waves.

## 6. Gyrogravitational Ratio

Gyrogravitational effect is defined to be the response of an angular momentum in a gravitomagnetic field produced by a gravitating source having a nonzero angular momentum. Ciufolini and E. C. Pavlis [117] have measured and verified this effect with 10-30 % accuracy for the dragging of the orbit plane (orbit angular momentum) of a satellite (LAGEOS) around a rotating planet (Earth) predicted for general relativity by Lense and Thirring [118]. Gravity Probe B [119] has measured and verified the dragging of spin angular momentum of a rotating quartz ball predicted by Schiff [120] for general relativity with 19 % accuracy. GP-B experiment has also verified the Second Weak Equivalence Principle (WEP II) for macroscopic rotating bodies to ultra-precision [121]. On 13 February 2012 the Italian Space Agency (ASI) launched the LARES (LAser RElativity Satellite) satellite with a Vega rocket for improving the measurement of Lense-Thirring effect together with other geodesy satellites [122]. On Earth, GINGER (Gyroscopes IN General Relativity) is a multi-ring-laser array project aimed to measure the Lense-Thirring effect to 1 % [123].

Just as in electromagnetism, we can define gyrogravitational factor as the gravitomagnetic moment (response) divided by angular momentum for gravitational interaction. We use macroscopic (spin) angular momentum in GR as standard, its gyrogravitational ratio is 1 by definition. In Ref. [124], we use coordinate transformations among reference frames to study and to understand the Lense-Thirring effect of a Dirac particle. For a Dirac particle, the wave-function transformation operator from an inertial frame to a moving accelerated frame is obtained. According to equivalence principle, this gives the gravitational coupling to a Dirac particle. From this, the Dirac wave function is solved and its change of polarization gives the gyrogravitational ratio 1 from the first-order gravitational effects. In a series of papers on spin-gravity interactions and equivalence principle, Obukhov, Silenko and Teryaev [125] have calculated directly the response of the spin of a Dirac particle in gravitomagnetic field and showed that it is the same as the response of a macroscopic spin angular momentum in general relativity (See, also, Tseng [126]). Randono have showed that the active frame-dragging of a polarized Dirac particle is the same as that of a macroscopic body with equal angular momentum [127]. All these results are consistent with EEP and the principle of action-equal-to-reaction. However, these findings do not preclude that the gyrogravitational ratio to be different from 1 in various different theories of gravity, notably torsion theories and Poincaré gauge theories.

What would be the gyrogravitational ratios of actual elementary particles? If they differ from one, they will definitely reveal some inner gravitational structures of elementary particles, just as different gyromagnetic ratios reveal inner electromagnetic structures of elementary particles. These findings would then give clues to the microscopic origin of gravity.

Promising methods to measure particle gyrogravitational ratio include [61]: (i) using spin-polarized bodies (e.g. polarized solid $He^3$, Dy-Fe, Ho-Fe, or other compounds) instead of rotating gyros in a GP-B type experiment to measure the gyrogravitational ratio of various substances; (ii) atom interferometry; (iii) nuclear spin gyroscopy; (iv) superfluid $He^3$ gyrometry. Notably, there have been great developments in atom interferometry [128,129] and nuclear gyroscopy [130]. However, to measure particle gyrogravitational ratios the precision is still short by several orders and more developments are required.

## 7. An Update of Search for Long Range/Intermediate Range Spin-Spin, Spin-Monopole and Spin-Cosmos Interactions

In this section, we update our review [61,131] on the search for the long range/intermediate range spin-spin, spin-monopole and spin-cosmic interactions.

*Spin-spin experiments*

Geomagnetic field induces electron polarization within the Earth. Hunter et al. [132] estimated that there are on the order of $10^{42}$ polarized electrons in the Earth compared to $\sim 10^{25}$ polarized electrons in a typical laboratory. For spin-spin interaction, from their results there is an improvement in constraining the coupling strength of the intermediate vector boson in the range greater than about 1 km [132].

*Spin-monopole Experiments*

In [61], we have used axion-like interaction Hamiltonian

$$H_{int} = [\hbar(g_s g_p)/8\pi mc]\,(1/\lambda r + 1/r^2)\exp(-r/\lambda)\,\boldsymbol{\sigma}\cdot\hat{\mathbf{r}}, \qquad (122)$$

to discuss the experimental constraints on the dimensionless coupling $g_s g_p/\hbar c$ between polarized (electron) and unpolarized (nucleon) particles. In (24), $\lambda$ is the range of the interaction, $g_s$ and $g_p$ are the coupling constants of vertices at the polarized and unpolarized particles, $m$ is the mass of the polarized particle and $\boldsymbol{\sigma}$ is Pauli matrix 3-vector. Hoedl *et al.* [133] have pushed the constraint to shorter range by about one order of magnitude since our last review [61]. In this update, we see also good progress in the measurement of spin-monopole coupling between polarized neutrons and unpolarized nucleons [134-136]. Tullney *et al.* [136] obtained the best limit on this coupling for force ranges between $3 \times 10^{-4}$ m and 0.1 m. Regards to a recent analysis of a direct spin-axion momentum interaction and its empirical constraints, please see Stadnik and Glambaum [137].

*Spin-cosmos experiments*

For the analysis of spin-cosmos experiments for elementary particles, one usually uses the following Hamiltonian:

$$H_{cosmic} = C_1\sigma_1 + C_2\sigma_2 + C_3\sigma_3, \qquad (25)$$

in the cosmic frame of reference for spin half particle with $C$'s constants and $\sigma$'s the Pauli spin matrices (see, e.g. [138] or [61]). The best constraint now is on bound neutron from a free-spin-precession $^3$He-$^{129}$Xe comagnetometer experiment performed by Allmendinger *et al.* [130]. The experiment measured the free precession of nuclear spin polarized $^3$He and $^{129}$Xe atoms in a homogeneous magnetic guiding field of about 400 nT. As the laboratory rotates with respect to distant stars, Allmendinger *et al.* looked for a sidereal modulation of the Larmor frequencies of the collocated spin samples due to (25) and obtained an upper limit of $8.4 \times 10^{-34}$ GeV (68% C.L.) on the equatorial component $C^n_\perp$ for neutron. This constraint is more stringent by $3.7 \times 10^4$ fold than the limit on that for electron [139]. Using a $^3$He-K co-magnetometer, Brown *et al.* [140] constrained $C^p_\perp$ for the proton to be less than $6 \times 10^{-32}$ GeV. Recently Stadnika and Flambaum [141] analyzed the nuclear spin contents of $^3$He and $^{129}$Xe together with a re-analysis of the data of Ref. [130] to give the following improved limit on $C^p_\perp$: $C^p_\perp < 7.6 \times 10^{-33}$ GeV.

## 8. Prospects

After the cosmological electoweak (vacuum) phase transition around 100 ps from the Big Bang, high energy photons came out. At this time it is difficult to do measurement, although things may still evolve according to precise physical law – notably quantum electrodynamics and classical electrodynamics. When our Universe cooled down, precision metrology became possible. Metrological standards could be defined and implemented according to the fundamental physical laws. The cosmic propagation according to Galileo's Weak Equivalence Principle for photons (nonbirefringence) in the framework of premetric classical electrodynamics of continuous media dictates that the spacetime constitutive tensor must be of core metric form with an axion (pseudoscalar) degree of freedom and a dilaton (scalar) degree of freedom. Propagation of pulsar pulses, radio galaxy signals and cosmological gamma ray bursts has verified this conclusion empirically down to $10^{-38}$, i.e. to $10^{-4} \times O([M_{Higgs}/M_{Planck}]^2)$. This is also the order that the generalized closure relations of electrodynamics are verified empirically. The axion and dilaton degrees of freedom are further constrained empirically in the present phase of the cosmos (Table 1). However, we should give a different thought to the axion and dilaton degrees of freedom in exploring spacetime and gravitation in the very early universe within 100 ps from the 'Big Bang'; we may need to look for imprints of new physics and new principles.

On the other hand, experiments with spin are important in verifying Galileo Equivalence Principle and Einstein Equivalence Principle which are important cornerstones of spacetime structure and gravitation. It is not surprising that cosmological observations on polarization phenomena become the ultimate test ground of the equivalence principles, especially for the photon sector. Some of the dispersion relation tests are reaching second order in the ratio of Higgs boson mass and Planck mass. Ultra-precise laboratory experiments are reaching ground in advancing constraints on various (semi-)long-range spin interactions. Sooner or later, experimental efforts will reach the precision of measuring the gyrogravitational ratios of elementary particles. All these developments may facilitate ways to explore the origins of gravity.

**Acknowledgements**

I would like to thank Sperello di Serego Alighieri for helpful comments on the manuscript.

**References**

[1] W.-T. Ni, Genesis of general relativity: a concise exposition, Chapter 2 in *One Hundred Years of General Relativity: from Genesis and Foundations to Gravitational Waves, Cosmology and Quantum Gravity*, edited by W.-T. Ni (World Scientific, Singapore, 2015).
[2] J. D. Jackson, *Classical Electrodynamics*, 2nd edition (Wiley, Hoboken, 1975)
[3] H. Minkowski, Die Grundgleichungen für die elektromagnetischen Vorgänge in bewegten Körpern, *Königliche Gesellschaft der Wissenschaften zu Göttingen. Mathematisch-Physikalische Klasse. Nachrichten,* pp. 53-111 (1908); this paper was read before the Academy on 21 December 1907; (English translation) The fundamental equations for electromagnetic processes in Moving bodies, translated from German by Meghnad Saha and Wikisource,

http://en.wikisource.org/wiki/Translation:The_Fundamental_Equations_for_Electro…

[4] See also F. W. Hehl, *Ann. Phys. (Berlin)* **17**, 691 (2008) for a historical account and detailed explanation.

[5] H. Bateman, The transformation of the electrodynamical equations, *Proc. Camb. Math. Soc., Ser. 2*, **8**, 223 (1909); received March 7, 1909, read March 11, 1909, revised July 22, 1909.

[6] F. W. Hehl and Yu. N. Obukhov, *Foundations of Classical Electrodynamics: Charge, Flux, and Metric* (Birkhäuser: Boston, MA, 2003).

[7] I. E. Tamm, *Electrodynamics of an anisotropic medium in special relativity theory*, Zhurn. Ross. Fiz.-Khim. Ob. **56**, n. 2‒3 (1924) 248‒262 (in Russian); Reprinted in: I. E. Tamm, *Collected Papers* (Nauka: Moscow, 1975) Vol. 1, pp. 19‒32 (in Russian).

[8] L. Mandelstam and J. Tamm, *Elektrodynamik der anisotropen Medien in der speziellen Relativitätstheorie*, Mathematische Annalen **95** (1926) 154‒160 [Errata ibid. **96** (1927) 600]; Reprinted in: I. E. Tamm, *Collected Papers* (Nauka: Moscow, 1975) Vol. 1, pp. 62‒67 (in Russian).

[9] M. v. Laue, *Die Relativitätstheorie, Vol. 1: Die spezielle Relativitätstheorie*, 5th rev. edition (Vieweg: Braunschweig, 1952).

[10] E. J. Post, *Formal Structure of Electromagnetics – General Covariance and Electromagnetics* (North Holland: Amsterdam, 1962; and Dover: Mineola, New York, 1997).

[11] A. Einstein and M. Grossmann, Entwurf einer ver-allgemeinerten relativitätstheorie und einer theorie der gravitation, Zeit. Math. Phys. 63, 215-225 (1913); See also, Outline of a generalized theory of relativity and of a theory of gravitation, in *The Collected Papers of Albert Einstein*, Vol. 4.

[12] A. Einstein, Die formale Grundlage der allgemeinen Relätivitatstheorie, Königlich Preußische Akademie der Wissenschaften (Berlin). Sitzungsberichte, 1914; See also, The formal foundation of the general theory of relativity, in *The Collected Papers of Albert Einstein*, Vol. 4.

[13] A. Einstein, Eine Neue Formale Deutung der Maxwellschen Feldgleichungen der Elektrodynamik, *Königlich Preußische Akademie der Wissenschaften* (Berlin), 184-188 (1916); See also, A new formal interpretation of Maxwell's field equations of Electrodynamics, in *The Collected Papers of Albert Einstein*, Vol. 6,

[14] W.-T. Ni, Equivalence Principles and Precision Experiments, in *Precision Measurement and Fundamental Constants II*, ed. by B. N. Taylor and W. D. Phillips, Natl. Bur. Stand. (U S) Spec. Publ. 617 (1984) 647.

[15] W.-T. Ni, Timing Observations of the Pulsar Propagations in the Galactic Gravitational Field as Precision Tests of the Einstein Equivalence Principle, in *Proceedings of the Second Asian-Pacific Regional Meeting of the International Astronomical Union on Astronomy, Bandung, Indonesia – 24 to29 August 1981*, ed. by B. Hidayat and M. W. Feast (Published by Tira Pustaka, Jakarta, Indonesia, 1984) pp. 441-448.

[16] W.-T. Ni, Equivalence Principles, Their Empirical Foundations, and the Role of Precision Experiments to Test Them, in *Proceedings of the 1983 International School and Symposium on Precision Measurement and Gravity Experiment*, Taipei, Republic of China, January 24-February 2, 1983, ed. by W.-T. Ni (Published by National Tsing Hua University, Hsinchu, Taiwan, Republic of China, 1983) pp. 491-517 [http://astrod.wikispaces.com/].

[17] W.-T. Ni, Spacetime structure and asymmetric metric from the premetric formulation of electromagnetism, *Phys. Lett. A*, http://dx.doi.org/10.1016/j.physleta.2015.03.004 (2015), arXiv:1411.0460.

[18] 1. I. Newton, *Philosophiae Naturalis Principia Mathematica* (London, 1687).

[19] G. Galilei, *Discorsi e dimostriazioni matematiche intorno a due nuove scienze* (Elzevir, Leiden, 1638). English translation by H. Crew and A. de Salvio, Dialogues Concerning Two New Sciences, Macmillan, New York, 1914; reprinted by Dover, New York, 1954.

[20] W.-T. Ni, *Phys. Rev. Lett.* **38** (1977) 301.

[21] W.-T. Ni, *Bull. Am. Phys. Soc.* **19** (1974) 655.

[22] A. Einstein, Ist die Trägheit eines Körpers won seinem Energieinhalt abhängig? *Ann. d. Phys.* **18**, 639 (1905).

[23] R. V. Eötvös, *Math. Naturwiss. Ber. Ungarn* **8**, 65 (1889).

[24] M. Planck, Berl. Sitz. 13 June 1907, p.542, specially at p.544
[25] A. Einstein, *Jahrb. Radioakt. Elektronik* **4**, 411 (1907); Corrections by Einstein in *Jahrb. Radioakt. Elektronik* **5**, 98 (1908); English translations by H. M. Schwartz in *Am. J. Phys.* **45**, 512, 811, 899 (1977).
[26] C.W. Misner, K. S. Thorne, and J. A. Wheeler, *Gravitation* (Freeman, 1973).
[27] A.S. Eddington, A generalization of Weyl's theory of the electromagnetic and gravitational fields. *Proc. R. Soc. Lond.* **A99** 104 (1921).
[28] É. Cartan, Sur une généralisation de la notion de courbure de Riemann et les espaces à torsion. (27.5. Acade. Sci. (Paris) **174**, 593 (1922).
[29] É. Cartan, Sur les variétés à Connexion affine et la théorie de la relativitée généralisée I, I (suite), II. Ann. Ec. Norm. Sup. 40 (1923), 325; 41 (1924), 1; 42 (1925), 17.
[30] O. Stern, Zeit. f. Phys. l, 249 (1921); O. Stern and W. Gerlach, Zeit. f. Phys. 8, 110; 9, 349 (1922).
[31] G. Uhlenbeck and S. Goudsmit, Naturwiss. 13, 953 (1925); Nature 117, 264 (1926).
[32] D.W. Sciama, On the analogy between charge and spin in general relativity, in Recent Developments in General Relativity (Pergamon + PWN, Oxford, 1962), p.415.
[33] D.W. Sciama, The physical structure of general relativity. Rev. Mod. Phys. 36, 463 and 1103 (1964).
[34] T.W.B. Kibble, Lorentz invariance and the gravitational field. J. Math. Phys. 2 212 (1961).
[35] R. Utiyama, Phys. Rev. 101, 1579 (1956).
[36] F. W. Hehl, J. Nitsch, and P. von der Heyde, Poincaré gauge field theory with quadratic Lagrangian, in A. Held (ed.), *General Relativity and Gravitation – One Hundred Years after the Birth of Albert Einstein, vol. 1* (Plenum, New York, 1980), pp.329-355.
[37] K. Hayashi and T. Shirafujii, Gravity from Poincaré gauge field theory of fundamental particles. I, *Prog. Theor. Phys.* 61, 866-882 (1980).
[38] F. W. Hehl, P. von der Heyde, G. D. Kerlick, and J. M. Nester, *Rev. Mod. Phys.* **48**, 393 (1976).
[39] P. von der Heyde, *Nuovo Címento Lett.* **14**, 250 (1975).
[40] W.-T. Ni, Spin, Torsion and Polarized Test-Body Experiments, in *Proceedings of the 1983 International School and Symposium on Precision Measurement and Gravity Experiment*, Taipei, Republic of China, January 24-February 2, 1983, ed. by W.-T. Ni (Published by National Tsing Hua University, Hsinchu, Taiwan, Republic of China, 1983) pp. 532-540 [http://astrod.wikispaces.com/].
[41] W.-T. Ni, *Phys. Lett. A* **120**, 174-178 (1986).
[42] K. Nordtvedt, Jr., *Phys. Rev.* **169**, 1014, 1017 (1968), and **170**, 1186 (1968).
[43] R. H. Dicke, *Gravitation and the Universe* (American Philosophical Society. Philadelphia, Pa, 1969), p. 19-24.
[44] R. H. Dicke, lectures in *Relativity, Groups, and Topology*, C. and B. DeWitt (Gordon and Breach, New York, 1964).
[45] C. M. Will and K. Nordtvedt, Jr., *Astrophys. J.* **77**, 757 (1972).
[46] K. Nordtvedt, Jr., and C. M. Will, *Astrophys. J.* **77**, 775 (1972).
[47] A. M. Nobili et al., *Amer. J. Phys.* **81**, 527 (2013).
[48] E. Di Casola, S. Liberati, and S. Sonego, *Amer. J. Phys.* **83**, 39-46 (2015).
[49] L. I. Schiff, *Am. J. Phys.* **28**, 340 (1960).
[50] R .H. Dicke, *Am. J. Phys.* **28**, 344 (1960).
[51] K. S. Thorne, D. L. Lee, and A. P. Lightman, *Phys. Rev. D* **7,** 3563 (1973).
[52] A.P. Lightman, and D.L. Lee, *Phys. Rev. D* **8**, 364 (1973).
[53] W.-T. Ni, A Nonmetric Theory of Gravity, preprint, Montana State University, Bozeman, Montana, USA (1973) [http://astrod.wikispaces.com/].
[54] P. Wolf *et al.*, *Nature* **467** (2010) E1.
[55] H. Müller, A. Peters and S. Chu, *Nature* **467** (2010) E2.
[56] J. M. Gérard, Class. Quantum Grav. **24**, 1867 (2007).
[57] E. Di Casola, S. Liberati, and S. Sonego, *Phys. Rev. D* **89,** 084053 (2014).
[58] M. Haugan and T. Kauffmann, Phys. Rev. D **52**, 3168 (1995).
[59] C. Lämmerzahl and F. W. Hehl, *Phys. Rev. D* **70**, 105022 (2004).

[60] W.-T. Ni, *Prog. Theor. Phys. Suppl.* **172**, 49 (2008) [arXiv:0712.4082].
[61] W.-T. Ni, *Reports on Progress in Physics* **73**, 056901 (2010).
[62] W.-T. Ni, *Phys. Lett. A* **378**, 1217-1223 (2014).
[63] W.-T. Ni, *Phys. Lett. A* **378**, 3413 (2014).
[64] A. Favaro and L. Bergamin, *Annalen der Physik* **523**, 383-401 (2011).
[65] M. F. Dahl, *Journal of Physics A: Mathematical and Theoretical* **45,** 405203 (2012).
[66] See e.g., F. G. Smith, Pulsars, Cambridge University Press (Cambridge, UK 1977).
[67] H.-W. Huang, Pulsar timing and equivalence principle tests, *Master Thesis*, National Tsing Hua University (2002).
[68] P.M. McCulloch, P.A. Hamilton, J.G. Ables, A.J. Hunt, *I.A.U. Circ.* (USA), No.3703, 1 (15 June 1982).
[69] D.C. Backer, S.R. Kulkarni, C. Heiles, M.M. Davis and W.M. Goss, *Nature* **300**, 615 (1982).
[70] V. A. Kostelecky and M. Mewes, *Phys. Rev. D* **66**, 056005 (2002).
[71] J. W. Moffat, in *Gravitation 1990 Proceedings of the Banff Summer Institute, Banff, Canada*, R. D. Mann and P. Wesson, eds. (1991).
[72] N. J. Cornish, J. W. Moffat and D. C. Tatarshi, *Gen. Rel. Grav.* **27,** 933-946 (1995).
[73] T. P. Krisher, *Phys. Rev. D* **44,** R2211 (1991).
[74] D. Götz, S. Covino, A. Fernández-Soto, P. Laurent, and Ž. Bosnjak, *Monthly Notice of Royal Astronomical Society* **431**, 3550 (2013).
[75] D. Götz et al., *Monthly Notice of Royal Astron. Soc.* **444**, 2776 (2014).
[76] P. Laurent, D. Götz, P. Binétruy, S. Covino, A. Fernández-Soto, *Phys. Rev. D* **83**, 12 (2011).
[77] V. A. Kostelecký and M. Mewes, *Phys. Rev. Lett.* **110**, 201601 (2013).
[78] W.-T. Ni, *Chin. Phys. Lett.* **22**, 33-35 (2005).
[79] Y. N. Obukhov, F. W. Hehl, *Phys. Lett. A* **341**, 357 (2005).
[80] Y. Itin, *Gen. Rel. Grav.* **40**, 1219 (2008).
[81] D. J. Fixsen, *Astrophys. J.* **707**, 916 (2009).
[82] S. di Serego Alighieri, W.-T. Ni, W.-P. Pan, *Astrophys. J.* **792**, 35 (2014).
[83] H.-H. Mei W.-T. Ni, W.-P. Pan, L. Xu, and S. di Serego Alighieri, *Astrophys. J.* **805**, 107 (2015).
[84] S. di Serego Alighieri, *Int. J. Mod. Phys. D* **24** (2015) 1530006, arXiv:1501.06460.
[85] P. A. R. Ade *et al*. *Astron. Astrophys.* 571 (2014) A16.
[86] Yu. N. Obukhov and F. W. Hehl, *Phys. Rev. D* **70**,125105 (2004).
[87] A. S. Eddington, *The mathematical theory of relativity*, 2nd edition (Cambridge Univ. Press, 1924).
[88] A. Einstein and E. G. Straus, *Ann. Math.* **47**, 731 (1946).
[89] E. Schrödinger, *Proc. R. Ir. Acad.* **51A**, 163 (1947).
[90] E. Schrödinger, *Space-time structure* (Cambridge University Press, 1950).
[91] I. V. Lindell and K. H. Wallén, *Journal of Electromagnetic Waves and Applications* **18** (2004) 957-968.
[92] A. Favaro, *Recent advances in classical electromagnetic theory*, PhD thesis, Imperial College London, 2012.
[93] R. Toupin, Elasticity and Electromagnetics, in *Non-linear continuum theories, C.I.M.E. Conference, Bressanone, Italy (1965)*, C. Truesdell and G. Grioli, Eds., pp. 203-342.
[94] M. Schönberg, Electromagnetism and Gravitation, *Revista Brasileira de Fisica* **1**, 91 (1971).
[95] A. Jadczyk, Electromagnetic permeability of the vacuum and light-cone structure, *Bulletin de l'Academie Polonaise des Sciences -- Séries des sciences physiques et astron.* **27**, 91 (1979).
[96] R. D. Peccei and H. R. Quinn, *Phys. Rev. Lett.* **38**, 1440 (1977).
[97] S. Weinberg, *Phys. Rev. Lett.* **40**, 233 (1978).
[98] F. Wilczek, *Phys. Rev. Lett.* **40**, 279 (1978).
[99] J. Kim, *Phys. Rev. Lett.* **43**, 103 (1979).
[100] M. A. Shifman, A. I. Vainshtein and V. I. Zakharov, *Nucl. Phys. B* **166,** 493 (1980)
[101] M. Dine, Fischler and M. Srednicki, *Phys. Lett.* **104B**, 199 (1981).
[102] S.-L. Cheng, C.-Q. Geng and W.-T. Ni, *Phys. Rev. D* **52,** 3132 (1995) and references therein.
[103] M. Yu. Khlopov, *Cosmoparticle physics* (World Scientific, 1999); and references therein.
[104] W.-T. Ni, *Phys. Lett. A* **120,** 174 (1987).

[105] W.-T. Ni, Some Basic Points about Metrology, in *Proceedings of the 1983 International School and Symposium on Precision Measurement and Gravity Experiment*, Taipei, Republic of China, January 24-February 2, 1983, ed. by W.-T. Ni (Published by National Tsing Hua University, Hsinchu, Taiwan, Republic of China, 1983) pp. 121-134 [http://astrod.wikispaces.com/].
[106] B. W. Petley, Fundamental Physical Constants and the Frontier of Measurement (Bristol [Avon]; Boston: A. Hilger, 1985).
[107] T. J. Quinn, *IEEE Trans. Instrum. Meas.* **40**, 81 (1991).
[108] P. Becker, *Contempory Physics* **53**, 461 (2012).
[109] C. A. Sanchez, B. M. Wood, R. G. Green, J. O. Liard and D. Inglis, *Metrologia* **51**, S5-S14 (2014).
[110] D. B. Newell, *Phys. Today*, July 2014, 35 (2014).
[111] R. D. Vocke, S. A. Raab and G. C. Turk, *Metrologia* **51**, 361-375 (2014).
[112] B. Andreas *et al*., *Metrologia* **48**, S1-13 (2011).
[113] L. Yang, Z. Mester, R. E. Sturgeon and Meija, *Anal. Chem.* **84**, 2321-7 (2012).
[114] P. J. Mohr, B. N. Taylor, D. B. Newell, Rev. Mod. Phys. 84, 1527-1605 (2012)
[115] A. D. Ludlow, M. M. Boyd, J. Ye, E. Peik, and P. O. Schmidt, *Rev. Mod. Phys.* **87**, 637 (2015).
[116] K. Kuroda, W.-T. Ni, and W.-P. Pan, Gravitational waves: classification, methods of detection, sensitivities, and sources, Chapter 10 in *One Hundred Years of General Relativity: From Genesis and Empirical Foundations to Gravitational Waves, Cosmology and Quantum Gravity*, ed. W.-T. Ni (World Scientific, Singapore, 2015); *Int. J. Mod. Phys. D* **24**, 1530031 (2015).
[117] I. Ciufolini and E. C. Pavlis, *Nature* **431**, 958 (2004).
[118] J. Lense and H. Thirring, *Phys. Z.* **19**, 156 (1918).
[119] C. W. F. Everitt, *et al.*, *Phys. Rev. Lett.,* **106**, 221101 (2011).
[120] L. I. Schiff, *Phys. Rev. Lett.* **4**, 215 (1960).
[121] W.-T. Ni, *Phys. Rev. Lett.,* **107,** 051103 (2011).
[122] I. Ciufolini, *et al.*, *Eur. Phys. J. Plus* **127**, 133 (2012).
[123] F. Bosi, G. Cella, A. Di Virgilio, *et al*., *Phys. Rev. D* **84**, 122002 (2011).
[124] Y.-C. Huang and W.-T. Ni, Propagation of Dirac Wave Functions in Accelerated Frames of Reference, arXiv:gr-qc/0407115.
[125] Y. N. Obukhov, A. J. Silenko, O. V. Teryaev, *Phys. Rev. D* **88**, 084014 (2013); and references therein.
[126] H.-H. Tseng, *On the Equation of Motion of a Dirac Particle in Gravitational Field and its Gyro-Gravitational Ratio*, M. S. (In Chinese with an English abstract, Advisor: W.-T. Ni), National Tsing Hua University, Hsinchu, 2001, for a derivation in the weak field limit.
[127] A. Randono, *Phys. Rev. D* **81**, 024027 (2010).
[128] T. Schuldt *et al*., *Exp. Astron.* **39**, 167 (2015).
[129] L. Zhou *et al*., *Phys. Rev. Lett.* **115**, 013004 (2015).
[130] F. Allmendinger *et al*., *Phys. Rev. Lett.* **112**, 110801(2014).
[131] W.-T. Ni, Searches for the role of spin and polarization in gravity: a five-year update, arXiv:1501.07696.
[132] L. Hunter *et al*., *Science* **339**, 928 (2013).
[133] S. A. Hoedl *et al*., *Phys. Rev. Lett.* **106**, 100801 (2011).
[134] P.-H. Chu *et al*., *Phys. Rev. D* **87**, 011105(R) (2013).
[135] M. Bulatowicz, *et al*., *Phys. Rev. Lett.* **111**, 102001 (2013).
[136] K. Tullney *et al*., *Phys. Rev. Lett.* **111**, 100801 (2013).
[137] Y. V. Stadnik and V. V. Flambaum, *Phys. Rev. D* **89**, 043522 (2014)
[138] P. R. Phillips, *Phys. Rev.* **139** B491-B494 (1965).
[139] B. R. Heckel, E. G. Adelberger, C. E. Cramer, T. S. Cook, S. Schlamminger and U. Schmidt, *Phys. Rev. D* **78,** 092006 (2008).
[140] J. M. Brown, S. J. Smullin, T. W. Kornack, and M.V. Romalis, *Phys. Rev. Lett.* **105**, 151604 (2010).
[141] Y. V. Stadnik and V. V. Flambaum, *Eur. Phys. J. C* **75**, 110 (2015).